\documentclass[aos,MSNbibl,seceqn,dvips]{arximspdf}
\usepackage{dcolumn}
\usepackage{graphicx}

\doi{10.1214/14-AOS1265} 
\volume{43}
\issue{1}
\pubyear{2015}
\firstpage{57}
\lastpage{89}
\docsubty{FLA}

\makeatletter
\newcolumntype{d}[1]{D{.}{.}{#1}}
\newcommand{\rrvert}{\vert}
\newcommand{\rrVert}{\Vert}
\newcommand{\llvert}{\vert}
\newcommand{\llVert}{\Vert}
\newtheorem{teo}{Theorem}[section] 
\newtheorem{lemma}{Lemma}[section] 
\newproclaim{rem}{Remark}
\newproclaim{ex}{Experiment}
\newcommand{\DPD}{\mathit{DPD}}
\newcommand{\DPDB}{\mathit{DPDB}}
\newcommand{\OSC}{\mathit{OSC}}
\newcommand{\err}{\mathrm{err}}
\renewcommand{\sc}{\mathrm{sc}}
\newcommand{\rep}{\mathit{rep}}
\newcommand{\margmin}{\operatorname{argmin}}
\newcommand{\diag}{\operatorname{diag}}
\newcommand{\hamm}{\mathrm{Hamm}}
\newcommand{\thmax}{\theta_{\max}}
\newcommand{\thmin}{\theta_{\min}}
\newcommand{\mumax}{\mu_{\max}}
\newcommand{\mumin}{\mu_{\min}}
\newcommand{\hlam}{\hat{\lambda}}
\newcommand{\heta}{\hat{\eta}}
\newcommand{\eigsp}{\mathrm{eigsp}}
\makeatother

\begin{document}
\begin{frontmatter}

\title{Fast community detection by SCORE}
\runtitle{Community detection by SCORE}

\begin{aug}
\author{\fnms{Jiashun}~\snm{Jin}\corref{}\thanksref{T1}\ead[label=e1]{jiashun@stat.cmu.edu}}
\runauthor{J. Jin}
\affiliation{Carnegie Mellon University}
\address{Department of Statistics\\
Carnegie Mellon University\\
Pittsburgh, Pennsylvania 15213\\
USA\\
\printead{e1}} 
\end{aug}
\thankstext{T1}{Supported in part by NSF Grant DMS-12-08315.}

\received{\smonth{8} \syear{2013}}
\revised{\smonth{8} \syear{2014}}

%
\begin{abstract}
Consider a network where the nodes split into $K$ different
communities. The community labels for the nodes are unknown and it is
of major interest to estimate them (i.e., community detection).
\textit{Degree Corrected Block Model} (DCBM) is a popular network
model. How to detect communities with the DCBM is an interesting problem,
where the main challenge lies in the degree heterogeneity.

We propose a new approach to community detection which we call the {S}pectral {C}lustering {O}n {R}atios-of-{E}igenvectors
(SCORE). Compared to classical spectral methods, the main innovation is
to use the entry-wise ratios between the first leading eigenvector and
each of the other leading eigenvectors for clustering. Let $A$ be the
adjacency matrix of the network.
We first obtain the $K$ leading eigenvectors of $A$, say, $\hat{\eta
}_1, \ldots, \hat{\eta}_K$, and let $\hat{R}$ be the $n \times
(K-1)$ matrix such that
$\hat{R}(i, k) = \hat{\eta}_{k+1}(i) / \hat{\eta}_1(i)$, $1 \leq i
\leq n$, $1 \leq k \leq K-1$. We then use $\hat{R}$ for clustering by
applying the $k$-means method.

The central surprise is, the effect of degree heterogeneity is largely
ancillary, and can be effectively removed by taking entry-wise ratios
between $\hat{\eta}_{k+1}$ and $\hat{\eta}_1$, $1 \leq k \leq K-1$.

The method is successfully applied to the web blogs data and the karate
club data, with error rates of $58/1222$ and $1/34$, respectively.
These results are
more satisfactory than those by the classical spectral methods.
Additionally, compared to modularity methods, SCORE is easier to
implement, computationally faster, and also has smaller error rates.

We develop a theoretic framework where we show that under mild conditions,
the SCORE stably yields consistent community detection.
In the core of the analysis is the recent development on Random Matrix
Theory (RMT),
where the matrix-form Bernstein inequality is especially helpful.
\end{abstract}

%
\end{frontmatter}

\setcounter{footnote}{1}

\section{Introduction} \label{secIntro}
Driven by the emergence of online ``networking communities'' (e.g.,
Facebook, LinkedIn, MySpace, Google$+$) and
by the growing recognition of scientifically central
networked phenomena (e.g., gene regulatory networks, citation networks,
road networks), we see today a great demand for methods to
infer the presence of network phenomena, particularly in the presence
of large datasets.
Tools and discoveries in this area could potentially reshape
scientific data analysis and even have impacts on daily life
(friendship, marketing, security).

A problem that is of major interest is ``network community detection''
\cite{BickelChen2009,FanLu,Choi,Hoff,Newman1,Newman2,Wolfe,Yu,Zhao,Zhaoetal}.
Given an $n$-node (undirected) graph ${\mathcal N} = (V, E)$, where $V
= \{1, 2, \ldots, n\}$
is the set of nodes and $E$ is the set of edges.
We believe that $V$
partitions into a small number of (disjoint) subsets or ``communities''.
The nodes within the same community
share some common characteristics. The community labels are unknown to us
and the main interest is to estimate them.

An iconic example is the web blogs data \cite{AdamicGlance2005}, which
was collected right after the 2004 presidential election. Each node of
the network is a web blogs about US politics, and each edge indicates a hyperlink
between them (we neglect the direction of the hyperlink
so that the graph is undirected).
In this network, there are two perceivable communities:
political liberal and political conservative.
It is believed that the web blogs share some common
political characteristics (liberal or conservative, one supposes)
that are significantly different between two communities, but
are not significantly different among the nodes in the same
community.

\subsection{Degree corrected block model (DCBM)}
In the spirit of ``all models are wrong, but some are useful'' \cite{Box},
we wish to find a network model that is both realistic
and mathematically tractable.

The stochastic block model (BM) is a classic network model. The BM
is mathematically simple and relatively easy to analyze \cite{BickelChen2009}. However, it is too restrictive to
reflect some prominent empirical characteristics
of real networks. For example, the BM implies
that the nodes within each community have
the same expected degrees, and that the distribution
of degrees within the community is Poisson.
However, this conflicts
with the empirical observation that in many natural networks, the degrees
follow approximately a power-law
distribution \cite{Survey,Kolaczyk}.

In a different line of development, there are the $p^*$ model and the
exponential random graph model (ERGM) \cite{Survey}. Compared to the BM,
these models are more flexible, but unfortunately are also more
complicated and so comparably much harder
to analyze.

DCBM is a recent model proposed by \cite{KarrerNewman2011},
which has become increasingly popular in network analysis \cite{BickelChen2009,FanLu,KarrerNewman2011,Yan,Zhaoetal}.
Compared to the BM, DCBM allows for degree heterogeneity and is much
more realistic: for each node, it uses a free parameter to model the degree.

The comparison of DCBM with the $p^*$ model and the ERGM \cite{Survey,Kolaczyk} is not obvious, given that all of them use a large number of
parameters.
However, in sections below, we propose a new spectral method where
we show that in the DCBM, the degree heterogeneity parameters are
largely ancillary: as far as community detection concerns, it is almost
unnecessary to estimate these heterogeneity parameters. For this
reason, the
DCBM is much easier to analyze than the $p^*$ or the ERGM model.

Perhaps the easiest way to describe the DCBM is to start with the case
of two communities (discussion on the case of $K$ communities is in
Section~\ref{secmain}).
Recall that ${\mathcal N} = (V, E)$ denotes an undirected network. We
suppose the nodes split into two (disjoint) communities as follows:
\[
V = V^{(1)} \cup V^{(2)}.
\]
Let $A$ be the $n \times n$ adjacency matrix of ${\mathcal N}$. In the
DCBM, we fix $(n + 3)$ positive parameters $(a, b, c)$ and $\{\theta
^{(n)}(i) \}_{i = 1}^n$ and assume that:
\begin{itemize}
\item$A$ is symmetric, with zeros on the diagonal (so there is no
self-connections).
\item The elements of the upper triangular $\{A(i,j)\dvtx  1 \leq i < j
\leq n \}$ are independent Bernoulli random variables satisfying
\[
P \bigl(A(i,j) = 1 \bigr) = \theta^{(n)}(i) \theta^{(n)}(j)
\cases{ a, &\quad$i, j \in V^{(1)}$,
\cr
c, &\quad$i, j \in
V^{(2)}$,
\cr
b, &\quad otherwise.}
\]
\end{itemize}
As $n$ changes, we assume $(a, b, c)$ are fixed but $\theta^{(n)}(i)$
may vary with $n$. The superscript ``$n$'' becomes tedious, so for
simplicity, we drop it from now on.
We call $\{\theta(i)\dvtx  1 \leq i \leq n\}$ the \textit{degree heterogeneity
parameters} or \textit{heterogeneity parameters} for short.

For identifiability, we assume
\[
\max\{a, b, c\} = 1, \qquad\thmax\leq g_0,
\]
where $\thmax= \max_{1 \leq i \leq n} \{\theta(i)\}$ and $g_0 \in
(0,1)$ is a constant.

It is probably more convenient if we rewrite the model in the matrix form.
The following notation is associated with the heterogeneity parameters
$\{\theta(i)\}_{i = 1}^n$ and are frequently used in this paper.
Let $\theta$ and $\Theta$ be the $n \times1$ vector and the $n
\times n$ diagonal matrix defined as follows:
%
\begin{equation}
\label{DefineTheta} \theta= \bigl(\theta(1), \theta(2), \ldots, \theta (n)
\bigr)', \qquad\Theta(i,i) = \theta(i), \qquad1 \leq i \leq n.
\end{equation}
Moreover, for $k = 1, 2$, let
${\mathbf1}_k$ be the $n \times1$ indicator vector such that ${\mathbf
1}_k(i) = 1$ if $i \in V^{(k)}$ and $0$ otherwise. With this notation,
we can rewrite
\[
A = E[A] + W, \qquad W \equiv A - E[A],
\]
where $E[A]$ denotes the expectation of $A$ (also an $n \times n$
matrix), and
\[
E[A] = \Omega- \diag(\Omega), \qquad\Omega\equiv\Theta \bigl[ a {
\mathbf1}_1 {\mathbf1}' + c {\mathbf1}_2 {
\mathbf1}_2' + b \bigl({\mathbf1}_1 {
\mathbf1}_2' + {\mathbf1}_2 {
\mathbf1}_1' \bigr) \bigr] \Theta.
\]
Note that the entries in the upper triangular of $W$ are independently
(but not identically) distributed as centered-Bernoulli; such $W$ is
known as a generalized Wigner matrix \cite{GW}.

\begin{rem*}
While it seems ${\mathbf1}_k$ are known, they are not for they
depend on $V^{(k)}$---the unknown community partitions that are of
primary interest.
\end{rem*}

\subsection{Where is the information: Spectral analysis heuristics}
In \cite{Tukey}, John Tukey mentioned an idea that can serve as a
general guideline for
statistical inference. Tukey's idea is that before we tackle any
statistical problem, we should think about ``which part of the data
contains the information'': the
``best'' procedure should capture the most direct information containing
the quantity of interest.

In our setting, the quantities of the interest are the community
labels. Recall that
\[
A = \Omega- \diag(\Omega) + W.
\]
Seemingly, $\Omega$ contains the most direct information of the
community labels: the matrix $W$ only contains noisy and indirect
information of the labels, and the matrix $\diag (\Omega)$
only has a negligible effect, compared to that of $\Omega$.

In light of this, we take a close look on $\Omega$. For $k = 1,2$, let
$\theta^{(k)} \equiv\theta^{(n,k)}$ be the $n \times1$ vector such
that (recall that $\theta$ is the shorthand for $\theta^{(n)}$)
\[
\theta^{(k)}(i) = \theta(i)\qquad\mbox{if }i \in V^{(k)}
\]
\mbox{and}
\[
\theta^{(k)}(i) = 0\qquad\mbox{otherwise}, 1 \leq i \leq n.
\]
For any vector $x$, let $\llVert x\rrVert $ denote the $\ell^2$-norm.
Write for short
\[
d_k = \bigl\llVert\theta^{(k)} \bigr\rrVert/\llVert \theta
\rrVert, \qquad k = 1, 2.
\]
Note that $\llVert \theta^{(k)} \rrVert $ can be interpreted as the
\textit{overall
degree intensities} of the $k$th community.

In most part of the paper, the eigenvalues of interest are simple
(i.e., algebraic multiplicity $1$ \cite{Horn}).
The following lemma is a special case of Lemma~\ref{lemmabasicalg1},
which is proved in the supplementary material \cite{score}, Appendix~\textup{C}
[$\Theta$ is the diagonal matrix in (\ref{DefineTheta})].

\begin{lemma} \label{lemmaL}
If $a c \neq b^2$, then $\Omega$ has two simple nonzero eigenvalues
\[
\tfrac{1}{2} \llVert\theta\rrVert^2 \Bigl( a
d_1^2 + c d_2^2 \pm\sqrt
{ \bigl(a d_1^2 - cd_2^2
\bigr)^2 + 4 b^2d_1^2
d_2^2} \Bigr),
\]
and the associated eigenvectors $\eta_1$ and $\eta_2$ (with possible
nonunit norms) are
\[
\Theta \Bigl(b d_2^2 \cdot{\mathbf1}_1 +
\tfrac{1}{2} \Bigl[ cd_2^2 - a d_1^2
\pm\sqrt{ \bigl(a d_1^2 - c d_2^2
\bigr)^2 + 4 b^2 d_1^2
d_2^2 } \Bigr] \cdot{ \mathbf1}_2 \Bigr).
\]
\end{lemma}

The key observation is as follows.
Let $r$ be the $n \times1$ vector of the coordinate-wise ratios
between $\eta_1$ and $\eta_2$ (up to normalizations)
\[
r(i) = \frac{\eta_2(i) /\llVert \eta_2 \rrVert } { \eta_1(i) /
\llVert \eta_1 \rrVert }, \qquad1 \leq i \leq n.
\]
Define the $n \times1$ vector $r_0$ by
%
\begin{equation}
\label{Definer0} r_0(i) = \cases{ 1, &\quad$i \in V^{(1)}$,
\vspace*{3pt}\cr
\displaystyle- \biggl(\frac{ a d_1^2 - c d_2^2 + \sqrt{ (a d_1^2 - c
d_2^2)^2 + 4
b^2 d_1 d_2} } { 2 b d_1 d_2} \biggr)^2, &\quad$i \in
V^{(2)}$.}\hspace*{-30pt}
\end{equation}
Then by Lemma~\ref{lemmaL} and basic algebra,
\[
r \propto r_0.
\]
We are now ready to answer Tukey's query on ``where is the
information'': the
sign vector of $r$ is the place that contains the most direct
information of the community labels.

The central surprise is that, as far as community detection concerns,
the \textit{heterogeneity parameters} $\{\theta(i)\}_{i = 1}^n$ are
largely ancillary, and their influence can
be largely removed by taking the coordinate-wise ratio of $\eta_1$ and
$\eta_2$ as above [though~$r$
still depends on $(\theta, n)$, but the dependence is only through the
overall degree intensities $d_1$ and $d_2$]. This allows us to
successfully extract the information containing the community labels
without any attempt to estimate the heterogeneity parameters.

Compared to approaches
where we attempt to estimate the heterogeneity parameters, our approach has
advantages. The reason is that many real-world networks (e.g., web
blogs network) are sparse in the sense that the degrees for many nodes
are small. If we try to
estimate the heterogeneity parameters of such nodes, we get relatively
large estimation errors which may propagate to subsequent studies.

\subsection{SCORE: A new approach to spectral community detection}
The\break above observations motivate the following procedure for community
detection, which we call {S}pectral {C}lustering {O}n {R}atios-of-{E}igenvectors (SCORE).
\begin{longlist}[(a)]
\item[(a)] Let $\hat{\eta}_1$ and $\hat{\eta}_2$ be the two
unit-norm eigenvectors of $A$ associated with the largest and the
second largest eigenvalues (in magnitude), respectively.

\item[(b)] Let $\hat{r}$ be the $n \times1$ vector of
coordinate-wise ratios: $\hat{r}(i) = \heta_2(i) / \hat{\eta
}_1(i)$, $1 \leq i \leq n$.

\item[(c)] Clustering the labels by applying the $k$-means method
to the vector $\hat{r}$, assuming there are $\leq2$ communities in total.
\end{longlist}
The key insight is that, under mild conditions, we expect to see that
\[
\hat{\eta}_1 \approx\eta_1/\llVert\eta\rrVert
_1, \qquad\heta_2 \approx\eta_2/\llVert
\eta_2 \rrVert,
\]
where $\eta_1$ and $\eta_2$ are the two eigenvectors of $\Omega$
as in Lemma~\ref{lemmaL}. Comparing with~(\ref{Definer0}), we expect
to have
\[
\hat{r} \approx r \propto r_0.
\]

In step~(c), we use the $k$-means method. Alternatively, we could use
the hierarchical clustering method \cite{Tibsbook}. For most of the
numeric study in this paper, we use the $k$-means package in MATLAB. In
comparison, the performance of the $k$-means method and the hierarchical
method are mostly similar, and that of the latter is slightly worse sometimes.

Note that since $\hat{r}$
is one-dimensional, both methods are equivalent to \textit{simple
thresholding}. That is,
for some threshold $t$, we classify a node $i$, $1 \leq i \leq n$, to
one community if $\hat{r}(i) > t$, and to the other community otherwise.
Seemingly, the simplest choice is $t = 0$. Alternatively, one could use
a recursive
algorithm to determine the threshold: (a) estimate the community labels
by applying the simple thresholding to $\hat{r}$ with $t = 0$, (b)
update the threshold with the estimated labels, say, following (\ref
{Definer0}) with $(a, b, c, d_1, d_2)$ estimated, (c) repeat (a)--(b)
with the threshold updated recursively.

The computational complexity of SCORE mostly comes from
obtaining the two leading eigenvectors.
For many social network data sets, the adjacency matrix
is very sparse, and the computational complexity for obtaining
the leading eigenvectors is only slightly larger than $O(n^2)$, using
the simple power method. See \cite{Newman2}, page~8581, for more discussions.
\subsection{Consistency of SCORE}
\label{subsecconsi}
In Section~\ref{secmain}, we extend SCORE to the
case where we have $K$ communities ($K \geq2$), and investigate
the theoretic properties.
The main results are presented in Theorems~\ref{teoR} and~\ref{teomain}. In the case of $K = 2$, these theorems simplify to that
of tight bounds on $\llVert \hat{r} - r\rrVert $ and on the Hamming
error of
community detection (i.e., expected number of nodes where the estimated
label does not match with the true label).
A direct results of the two theorems is that, under some regularity conditions,
SCORE is (weakly) consistent \cite{Zhaoetal} in community detection,
in the sense that
the Hamming error is much smaller than $n$.

The focus of the study is to identify a class of $\theta$ that is
\textit{as broad as possible} over which SCORE is uniformly consistent.
To do so, we choose not to impose much \textit{structural} assumptions
on $\theta$ [such as an i.i.d. model for $\theta(i)$]. In fact, the regularity
conditions we need for the consistency are conditions that only depend
on the
\mbox{$\ell^q$-}norms and extreme coordinates of $\theta$, not on the
structure of $\theta$. The reasons for doing so is two-fold:
\begin{itemize}
\item The structure of $\theta$ is largely unknown. For example, the
correlation structures among different coordinates of $\theta$ is hard
to model and is hard to
estimate.
\item As far as community detection is concerned, the role of $\theta$
is largely
ancillary, and the effect of $\theta$ can be largely removed by using SCORE.
\end{itemize}
Note that our theoretic study is different from Zhao et al. \cite{Zhaoetal}
[where a scaled i.i.d. model is used for $\theta(i)$], and from \cite{Yu,Fishkind}
[where the focus is on BM so all $\theta(i)$ are equal].
On the other hand, compared to those in \cite{Zhaoetal,Yu,Fishkind},
since we choose not to impose much structural assumptions on $\theta$,
our results (and the regularity conditions) generally have
more complicate forms. To better compare our results with that in \cite{Zhaoetal,Yu,Fishkind}, we interpret our results and regularity
conditions in Section~\ref{subseciidmodel} with a (scaled) i.i.d. model.

\subsection{Applications to the web blogs data and the karate club data}
We investigate the performance of the SCORE with two well-known
networks: the
web blogs network and the karate club network. The web blogs network is
introduced earlier in the paper.
The network has a giant component which we use for the analysis. The
giant component consists of 1222 nodes and 16,714 edges. Each blog
is manually labeled either as liberal or conservative in \cite{AdamicGlance2005}
which we use as the ground truth.
The karate club network can be found in \cite{Zachary}. The network
consists of $34$ nodes and $136$ edges, where each node represents a
member in the club. Due to the fission of the club,
the network has two perceivable communities:
Mr. Hi's group and John's group. All members are labeled in \cite{Zachary}, Table~1, which we use as the ground truth.

\begin{figure}

\includegraphics{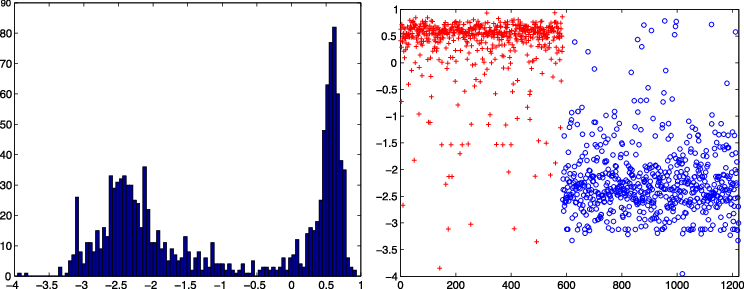}

\caption{The vector $\hat{r}$ (web blogs data). Left: histogram of
$\hat{r}$. Right: plot of the entries of $\hat{r}$ versus the node
indices (red cross: liberal; blue
circle: conservative).}\label{figblog1}
\end{figure}

Consider the web blogs network first.
In the left panel of Figure~\ref{figblog1},
we plot the histogram of the vector $\hat{r}$, which clearly shows a
two mode pattern, suggesting that there are two underlying communities.
In the right panel of Figure~\ref{figblog1},
we plot the entries of $\hat{r}$ versus the indices of the nodes, with
red crosses and blue circles
representing the nodes from the liberal and conservative communities,
respectively; the plot shows that the red crosses and blue circles are
almost completely separated from each other, suggesting that two
communities can be nicely separated by applying simple thresholding to
$\hat{r}$.

%

\begin{figure}[b]

\includegraphics{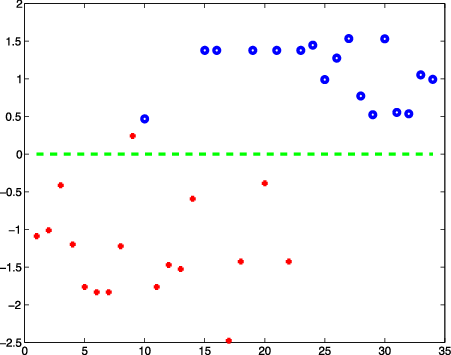}

\caption{Plot of the entries of $\hat{r}$ versus the node indices
(results are based on karate club network; red cross: Mr. Hi's group;
blue circle: John's group).}\label{figkarate1}
\end{figure}

The error rate of the SCORE is reasonably satisfactory. In fact,
if we use the procedure following steps (a)--(c), the error rate is $58/1222$.
The error rate stays the same if we replace the $k$-means method in (c)
by the hierarchical method (for both methods, we use the built-in
functions in MATLAB; the linkage for the hierarchical method is chosen
as ``average'' \cite{Tibsbook}).

Alternatively, we can use simple thresholding in step (c). In fact, the $k$-means
method is equivalent to simple thresholding with $t = -0.7$.
Moreover, the error rate is $82/1222$ if we set $t = 0$, and the error
rate is $55/1222$ if we set $t = -0.6$ (this is the ``ideal threshold,''
the threshold we would choose if we know the true labels; if only). The
results are tabulated in Table~\ref{tablerealdataerrors}, along with error
rates by some other methods, to be discussed below.

We consider the karate network next. Similarly, in Figure~\ref{figkarate1}, we plot the coordinates of $\hat{r}$ associated with
the karate data versus the node indices, with red crosses and blue
circles representing the nodes from the group of Mr. Hi and the group
of John \cite{Zachary}, respectively. Our method has an error rate of
$1/34$ if in step (c) we either
use the $k$-means method or the simple thresholding with $t = 0$ (the
error rate is $0/34$ if we set $t$ as the ``ideal threshold''). See
Table~\ref{tablerealdataerrors}
for details.

%

\subsection{Comparison with classical spectral clustering methods}
\label{subsecspectral}
The success of SCORE (which is a spectral method) prompts the question
whether classical spectral methods work well, too. Below are two
classical spectral methods:
\begin{longlist}[(a$'$)]
\item[(a$'$)] Obtain the two leading (unit-norm) eigenvectors $\hat
{\eta}_1$ and $\hat{\eta}_2$ of $A$.
\item[(b$'$)] Viewing $(\hat{\eta}_1, \hat{\eta}_2)$ as a
bivariate data set with sample size of $n$, apply the $k$-means method
assuming there are at most two communities.
\end{longlist}
Alternatively, one may use the following variation, which is studied in
\cite{Yu}.
\begin{longlist}[(a$''$)]
\item[(a$''$)] Obtain an $n \times n$ diagonal matrix $S$ by
$S(i,i) = \sum_{j = 1}^n A(i,j)$, $1 \leq i \leq n$.
\item[(b$''$)] Apply (a$'$)--(b$'$) to $S^{-1/2} A S^{-1/2}$.
\end{longlist}
We call the two procedures \textit{ordinary Principle Component Analysis}
(\textit{oPCA}) and \textit{normalized PCA} (\textit{nPCA}), respectively.\footnote{oPCA
and nPCA are also called spectral clustering on the adjacency matrix and
on the graph Laplacian, respectively. See \cite{FanLu,Chung}, for example.}

It turns out that both PCA approaches work unsatisfactorily.
In fact, for the web blogs data, the error rates of oPCA and nPCA are
$437/1222$ and $600/1222$, respectively, and for the karate data, the
error rates
are $1/34$ for both methods. See Table~\ref{tablerealdataerrors} for details.

\begin{figure}

\includegraphics{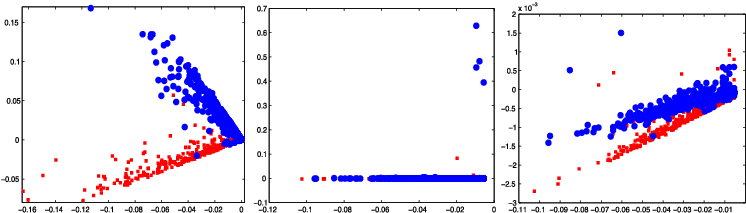}

\caption{Left: plot of the first leading eigenvector of $A$ ($x$-axis)
versus the second leading eigenvector of $A$ ($y$-axis). Middle: plot\vspace*{1pt}
of the first leading eigenvector of $S^{-1/2} A S^{-1/2}$ ($x$-axis)
versus the second leading eigenvector of $S^{-1/2} A S^{-1/2}$ ($y$-axis).
Right: zoom-in of the middle panel. Results are based on the web blogs data,
with red representing liberal and blue representing conservative.}\label{figblog2}
\end{figure}

The main reason why the two PCA methods perform unsatisfactorily is
that different coordinates of the two leading eigenvectors are heavily
affected by the degree inhomogeneity; see Lemma~\ref{lemmaL}. In the
left panel of Figure~\ref{figblog2},
we display the two leading eigenvectors of $A$, based on the web blogs
data. The coordinates of two vectors are highly skewed to the left,
reflecting serious degree heterogeneity. Compare \cite{eigenspoke}
where a similar phenomenon is observed.

%

Somewhat surprisingly, though nPCA intends to correct degree
heterogeneity, the correction is not particularly successful, partially
because that the adjacency matrix of the web blogs data is very sparse.
In the right two panels of Figure~\ref{figblog2} (the rightmost panel
is the zoom-in version of the panel to its left), we plot the two
leading eigenvectors of
$S^{-1/2} A S^{-1/2}$. It is seen that some of the entries of $\heta
_2$ are very large (compared to other entries).

Note\vspace*{1pt} that the unsatisfactory performance of oPCA (or nPCA)
does not mean that the two leading eigenvectors of $A$ (or $S^{-1/2} A
S^{-1/2}$) are not ``cluster-able''. It only means
that we need to pre-process the eigenvectors in a way
so that some conventional methods (such as the $k$-means) can cluster well;
SCORE provides a convenient pre-processing approach.

\subsection{Comparison with other spectral methods}
Newman \cite{Newman2} proposes a different spectral method, Spectral
Modularity (SM), which we have applied to the weblog data and the
karate data. The resultant error rates are $69/1222$ and $1/34$,
respectively, compared to $58/1222$ and $1/34$ by SCORE. In Section~\ref{secSimul}, we further compare this method with SCORE with
simulated data; see details therein.

Note that Newman's method is different from SCORE, especially when
there are $3$ or more communities. Note also that theoretically
Newman's method is not fully analyzed. In comparison, SCORE is fully
analyzed in Section~\ref{secmain},
where we discuss community detection for the general case of $K$ communities.

\subsection{Comparison with the profile likelihood approach}
The profile likelihood (PL) approach is a well-known method for
community detection \cite{BickelChen2009,KarrerNewman2011,Zhaoetal}.
The method was first proposed by Karrer and Newman
\cite{KarrerNewman2011} and was later carefully analyzed by Zhao et
al. \cite{Zhaoetal}; see \cite{Zhaoetal}, equation~(2.3), for details.


%
\begin{figure}

\includegraphics{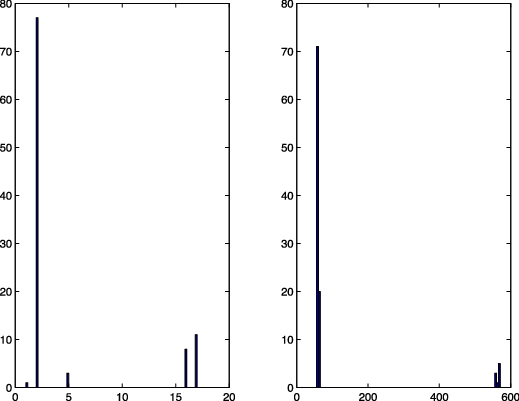}

\caption{Histogram of errors by PL for the karate data (left) and the
web blogs data. The results are based on $100$ independent repetitions.}\label{fighist1}
\end{figure}

In principle, PL is computationally NP-hard \cite{BickelChen2009} (and
so are many modularity methods; see, e.g., \cite{Zhaoetal}), as it
searches exhaustively over all possible community partitions, and pick
the one that optimizes the so-called functional of profile likelihood.
To mitigate this difficulty, many heuristic algorithms are proposed to
approximate the theoretic optimizer, among which is the so-called \textit{tabu algorithm}~\cite{Zhaoetal}.

We have compared SCORE with the PL (implemented with the tabu search;
the code is generously shared by authors of \cite{Zhaoetal}).
In comparison, PL is computationally much more expensive, and is
increasingly so when the size or complexity of the network increases.
The algorithm is also relatively unstable: it depends on the initial
guess of the community partition, so it may not converge to the true
partition with a ``bad'' starting point.
The instability can be alleviated by increasing the number of searches,
but that is at the expense of substantially longer computational time.

The error rates of PL for two data sets are illustrated in Figure~\ref{fighist1} (left: karate; right: web blogs), in terms of the
histograms based on $100$ independent repetitions
(the error rates are random for they depend on the initial guess of the
community partition, generated randomly).

The most prominent problem of PL (and many modularity methods \cite{Zhaoetal}) is that, in quite a few repetitions ($9$ out of $100$ for
the web blogs data, and $19$ out of $100$ for the karate data), the
algorithm fails to converge to the true community partition and yields
poor results. For the karate data, the number of clustering errors have
a mean of $4.85$ and a standard deviation of $5.7$. For the web blogs
data, the number of clustering errors have a mean of $104.5$ and a
standard deviation of $145.5$.
If we remove the ``outliers'' (the $9$ outlying repetitions for the web
blogs data and the $19$ outlying repetitions of the karate data), then
for the karate data, the errors have a mean of $2.1$ and a standard
deviation of $0.6$, and for the web blogs data, the mean is $59$, and
the standard deviation of $2.4$. See Table~\ref{tablerealdataerrors}.

That PL is more stable for the web blogs data than the karate data is
unexpected (as the karate data has a relatively small size, we expect
that it is relatively easy
for the PL to find the true community partition).
One possible explanation is that the communities in the former is more
strongly structured,
so the algorithm converges faster for the web blogs data than for the
karate data.

In the above experiments, we use a random start.
From time to time, one would like to first pick a fast algorithm to
estimate the labels, and then use the estimated labels to start the PL.
We have tried this approach where we start with the estimated labels
by oPCA, SCORE and nPCA. The error rates (based on $100$ independent
repetitions) for web blogs data are $62 \pm2.0$, $62 \pm0.0$ and
$569.4 \pm3.7$, correspondingly, and those for the karate data set are
$2 \pm0.0$ for all
three choices of start.

On one hand, this confirms that
PL performs well given a good start. On the other hand,
it is usually hard to pick a good start (or to evaluate how well a
start is) in practice, and the performance of
PL could be unsatisfactory given a poor start.

%
\begin{table}
\tabcolsep=0pt
\caption{Comparison of number errors. For SCORE, we consider three
different threshold choices.
The result of PL depend on the starting point and is random, where
means and standard deviations (SD) are computed based on $100$
independent repetitions}\label{tablerealdataerrors}
\begin{tabular*}{\tablewidth}{@{\extracolsep{\fill}}@{}ld{2.0}d{2.0}d{2.0}d{3.0}d{3.0}c@{}}
\hline
&\multicolumn{3}{c}{\textbf{SCORE}}& \multicolumn{2}{c}{\textbf{PCA}} & \\[-6pt]
&\multicolumn{3}{c}{\hrulefill}& \multicolumn{2}{c}{\hrulefill} &\\
&\multicolumn{1}{c}{$\bolds{t = 0}$} & \multicolumn{1}{c}{\textbf{$\bolds{k}$-means}} & \multicolumn{1}{c}{\textbf{Ideal}}
& \multicolumn{1}{c}{\textbf{Ordinary}} & \multicolumn{1}{c}{\textbf{Normalized}} & \multicolumn{1}{c@{}}{\textbf{PL}}
\\
\hline
Weblogs &82 &58 & 55 & 437 & 600 & 104.5 (SD:145.4)\\
Karate  &1  &1 & 0 & 1 & 1 & 4.9 (SD:5.7)\\
\hline
\end{tabular*}
\end{table}

To conclude this section, we mention some data analysis results (on one
or both data sets) in the literature, where the error rates are
reported in different forms.
The web blog data was analyzed in \cite{KarrerNewman2011}, where it
was reported that the normalized mutual information (NMI) between the
true labels and the estimated labels is $0.72$. In comparison, SCORE
yields an NMI of $0.725$. In \cite{Zhaoetal}, the error rate is
reported in terms of Adjusted Random Index (ARI) between the
true labels and the estimated labels. The ARI of SCORE is $0.819$ for
the web blog data and $0.8823$ for the karate data. The results are
similar to those reported in \cite{Chenetal}, page~16. The web blog data
is also analyzed in \cite{BickelChen2009}, with an error rate of $61/1228$.

\subsection{Summary}
We propose SCORE as a new approach to network community detection when a
DCBM is reasonable. The main innovation is to use the coordinate-wise
ratios of the leading eigenvectors for clustering. In doing so, we have
taken advantage of the fact that the degree heterogeneity parameters
$\theta(i)$ are merely nuisance and we can largely remove their
effects without actually estimating them.

We have used the karate club data and the web blogs data to investigate
the performances for several algorithms including SCORE, oPCA, nPCA,
Newman's SM and PL. First, SCORE behaves much more satisfactory than
the two PCA approaches. Second, SCORE is different from Newman's SM in
both the idea and in the algorithm (especially when $K > 2$), and has a
smaller error rate for analyzing the web blogs data. Third,
SCORE also has advantages over the PL: the good performance of PL
depends on a good start, so it can be unstable; also, computationally,
PL is comparably slower than SCORE, but it does not outperform in terms
of error rates. Finally, SCORE is conceptually simple and easy to
implement, so there is ample space for extensions in the future work.

The paper is closely related to \cite{Yu} (see also \cite{Choi}), but
is different in important ways. The focus of this paper is on DCBM
where the number of communities $K$ is small, while the focus of \cite{Yu} is on BM where $K$ is large. Our analysis
is also different from that in \cite{Zhaoetal}, for we use a
nonstructured model
for the degree heterogeneity parameters $\theta(i)$; see Sections~1.4
and 2
for more discussions.

\subsection{Content}
The remaining part of the paper is organized as follows. In Section~\ref{secmain}, we consider a $K$-community network with a fixed
integer $K \geq2$. By delicate
spectral analysis as in Sections~\ref{subsecSCORE}--\ref
{subsecHamm}, we lay out the framework under which the SCORE yields
consistent estimates of the community labels.
In Section~\ref{subsecstabi}, we address the stability of the SCORE,
where Lemmas~\ref{lemmaell2norm1}--\ref{lemmaell2norm2}
contain key ingredient for proving the main theorems. In
Section~\ref{subseciidmodel}, we compare our results with that in
\cite{Zhaoetal,Yu,Fishkind} using
a (scaled) i.i.d. model for $\theta(i)$.
In the supplementary material \cite{score}, Appendix~\textup{A}, we suggest some
extensions of the SCORE.
The main results are proved in the supplementary material \cite{score}, Appendix~\textup{B}.
where we outline main technical devices required for the proofs.
Numeric investigation is continued in Section~\ref{secSimul}, where we
compare SCORE, oPCA, nPCA, Newman's SM and PL with simulated data.
Section~\ref{secDiscu} discusses connection between SCORE and
existing literatures. Secondary lemmas are proved in the supplementary
material \cite{score}, Appendix~\textup{C}.
\subsection{Notation} \label{subsecnotation}
In this paper, for two vector $u, v$ with the same size, $(u, v)$
denotes their inner product. For any fixed $q > 0$ and any vector $x$,
$\llVert x\rrVert _q$ denotes the $\ell^q$-norm.
The subscript is dropped for simplicity if $q = 2$. For any matrix $M$,
$\llVert M\rrVert $ denotes
the spectral norm and $\llVert M\rrVert _F$ denotes the Frobenius
norm. For two
positive sequences $\{a_n\}_{n = 1}^{\infty}$ and $\{b_n\}_{n =
1}^{\infty}$, we say $a_n \sim b_n$ if $a_n/b_n \rightarrow1$ as $n
\rightarrow\infty$, and we say $a_n \asymp b_n$ if there is a
constant $c_0 > 1$ such that $1/c_0 \leq a_n / b_n \leq c_0$ for
sufficiently large $n$.

In this paper, the notation $\theta$ and $\Theta$ are always linked
to each other, where $\theta$ denotes the $n \times1$ vector of
degree heterogeneity parameters and $\Theta$ denotes the $n \times n$
diagonal matrix satisfying that $\Theta(i,i) = \theta(i)$, $1 \leq i
\leq n$.
Also, $\thmin= \min_{1 \leq i \leq n} \{ \theta(i) \}$ and $\thmax=
\max_{1 \leq i \leq n} \{ \theta(i)\}$. For a vector $\xi$, when all
coordinates are positive, we use $\OSC(\xi)$ to denote the oscillation
$\max_{1 \leq i, j \leq n} \{\xi(i)/\xi(j)\}$. Throughout the paper,
$C$ denotes a generic positive constant that may vary from occurrence
to occurrence.

\section{Main results} \label{secmain}

In this section, we consider the community detection problem where
the network ${\mathcal N} = (V, E)$ has $K$ communities. Throughout the
paper, $K \geq2$ is a known integer. See Section~\ref{secDiscu} for
discussion on the case where $K$ is unknown. The section contains the
main theoretic results of the paper, and is relatively long. Therefore,
it is necessary to give a
road map and to illustrate key ideas behind the main results.

First, in Section~\ref{subsecSCORE}, we extend the DCBM and SCORE
from the case of $K = 2$ to the case of $K \geq2$. We then carry out
spectral analysis on $\Omega$
and $A$, in Sections~\ref{subsecSCOmega} and~\ref{subsecSCX},
respectively.
In these sections, we derive explicit formulas for the leading
eigenvalues and leading
eigenvectors for $\Omega$ and $A$. The leading eigenvectors of $\Omega
$ and
$A$ (denoted by $\heta_k$ and $\eta_k$, $1 \leq k \leq K$,
resp.) have very similar formulas, where the differences are
bounded by terms
depending on the generalized Wigner matrix $W$. The study of these
terms boils down to that of controlling $\llVert \heta_k - \eta
_k\rrVert $ and $\llVert
\Theta^{-1}(\heta_k - \eta_k)\rrVert $. We present tight bounds on these
two quantities in
Lemmas~\ref{lemmaell2norm1}--\ref{lemmaell2norm2}. Such bounds
form the bases of Theorems~\ref{teoR}--\ref{teomain}, which are the
main result of the paper. Theorems~\ref{teoR} and~\ref{teomain}
are presented in Sections~\ref{subsecR} and~\ref{subsecHamm},
respectively.

The main regularity conditions we need are
(\ref{Asymp1}), (\ref{ConditionDAD})--(\ref{ConditionTheta}), and
(\ref{maintheta}), to be introduced.
Think the adjacency matrix $A$
as the sum of the ``signal matrix'' $E[A]$ and the ``noise matrix'' $A
- E[A]$.
Condition~(\ref{Asymp1}) ensures that the spectral norm of the noise
matrix is smaller than that of the signal matrix.
Conditions (\ref{ConditionDAD})--(\ref{ConditionTheta}) ensure that
$E[A]$ has well-spaced leading eigenvalues, so the associated leading
eigenvectors are robust to noise corruption. Condition (\ref
{maintheta}) ensures several sharp\vspace*{1pt} large-deviations inequalities for
vectors/matrices associated with $\llVert \heta_k - \eta_k\rrVert $
and $\llVert \Theta
^{-1}(\heta_k - \eta_k)\rrVert $ (some of these inequalities have only
become available very recently).

The goal of our analysis is different from that in \cite{Zhaoetal,Yu}; see Section~\ref{subsecconsi} for more discussion. The focus
here is to explore how broadly (in terms of $\theta$) SCORE is
consistent, so
we choose not to impose
much \textit{structural} assumptions on $\theta$.
Regularity conditions we impose only depend on the $\ell^q$-norms of
$\theta$ and
the extreme entries of $\theta$. As a result, some of these
conditions have relatively complicate form. In Section~\ref{subseciidmodel},
we revisit such conditions with a (scaled) i.i.d. model, and show that
these conditions can be reduced to simple forms, similar to those found
in the literature (e.g., \cite{Zhaoetal}).

\subsection{SCORE when there are $K$ communities}
\label{subsecSCORE}
Given an (undirected) network ${\mathcal N} = (V, E)$, we assume the
network splits into
$K$ different communities. That is,
the set of nodes $V$ partitions to $K$ different (disjoint) subsets:
\[
V = V^{(1)} \cup V^{(2)} \cup\cdots\cup V^{(K)}.
\]
Let $A$ be the adjacency matrix of ${\mathcal N}$, and introduce
%
\begin{equation}
\qquad {\mathcal M}_{n, K-1, K} = \bigl\{M\dvtx \mbox{$n \times(K-1)$ matrix that
has $ \leq K$ distinct rows} \bigr\}.
\end{equation}
SCORE for $K$-community networks contains the following steps (for
convenience, when we say ``leading eigenvectors'' or ``leading
eigenvalues'', we are comparing the \textit{magnitudes} of the eigenvalues,
neglecting the $\pm$1 signs):
\begin{itemize}
\item Obtain the $K$ (unit-norm) leading eigenvectors of $A$:
$\hat{\eta}_1, \heta_2, \ldots, \heta_K$.
\item Fixing a threshold $T_n$, define an $n \times(K-1)$ matrix $\hat
{R}^*$ such that for all $1 \leq i \leq n$ and $1 \leq k \leq K-1$,
%
\begin{eqnarray}\label{DefinehatR}
\hat{R}^*(i, k) = \cases{ \hat{R}(i,k), &\quad if $ \bigl\llvert
\hat{R}(i,k) \bigr\rrvert\leq T_n$,
\cr
T_n, &\quad if $
\hat{R}(i,k) > T_n$,
\cr
- T_n, &\quad if $\hat{R}(i,k) <
-T_n$}
\nonumber\\[-8pt]\\[-8pt]
\eqntext{\displaystyle\mbox{where } \hat{R}(i,k) = \frac{ \heta
_{k+1}(i)}{\heta_1(i)}.}
\end{eqnarray}
\item Let $M^*$ be the matrix satisfying
\[
M^* = \mathop{\margmin}_{M \in{\mathcal M}_{n, K-1, K} } \bigl\llVert\hat{R}^* - M \bigr\rrVert
_F^2.
\]
Write\vspace*{1pt} $M^* = [m_1, m_2, \ldots, m_n]'$ so that $m_i'$ is the $i$th row
of $M^*$.
Note that $M^*$ has at most $K$ distinct rows, say, $m_{i_1}',
m_{i_2}', \ldots, m_{i_K}'$ for some indices $1 \leq i_1 < \cdots<
i_K \leq n$.
We\vspace*{1pt} partition all nodes into $K$ communities
$\hat{V}^{(1)}, \hat{V}^{(2)}, \ldots, \hat{V}^{(K)}$ such that
$\hat{V}^{(k)} = \{1 \leq j \leq n\dvtx  m_j = m_{i_k} \}$.
\end{itemize}
Note that the last step is the classical $k$-means method.
We make the following remarks. First, when $M^*$ has $k$ distinct rows
for some $k < K$, we let $\hat{V}^{(\ell)} = \varnothing$ for all $(k
+ 1) \leq\ell\leq K$. Second, the choice of the threshold $T_n$ is
flexible, and for convenience, we take
%
\begin{equation}
\label{Definetn} T_n = \log(n)
\end{equation}
in this paper. We impose thresholding in (\ref{DefinehatR}) mainly for
technical convenience in the proof\vspace*{1.5pt} of Theorem~\ref{teomain}. Numeric
study in this paper suggests that
no coordinate of $\hat{R}$ would be unduly large and so the
thresholding procedure in (\ref{DefinehatR}) is rarely necessary. Last,
since $\heta_k$ are real-valued unit-norm eigenvectors, $1 \leq k \leq K$,
so by basic algebra, provided that the largest $K$ eigenvalues\vspace*{1pt} are all simple,
$\heta_k$ are uniquely determined, up to a factor of $\pm$1. Correspondingly,
all columns of $\hat{R}$ are uniquely determined, up to a factor of
$\pm$1; these factors do not affect the clustering results.

\begin{rem*}
In SCORE, we apply the $k$-means algorithm to the $n \times(K-1)$
matrix $\hat{R}^*$. When $K = 2$, the algorithm reduces to simple
thresholding so the computational cost is relatively low. For larger $K$,
the algorithm is NP hard. For numeric study, a conventional approach is
to use some heuristic methods. In this paper, we use the build-in
$k$-means package in MATLAB, which is one of such heuristic methods.
\end{rem*}

\subsection{DCBM when there are $K$ communities}
As before, we assume that the
adjacency matrix $A$ satisfies
%
\begin{equation}
\label{DefineX} A = E[A] + W, \qquad E[A] = \Omega- \diag (\Omega),
\end{equation}
where $\Omega$ is symmetric, and $W \equiv A - E[A]$ is the
generalized Wigner matrix.
In the core of DCBM is a $K \times K$ matrix
\[
P = \bigl( P(i,j) \bigr)_{1 \leq i, j \leq K}.
\]
For positive parameters $\{\theta(i)\}_{i = 1}^n$ as before, we extend the
$n \times n$ matrix $\Omega$ to a more general form such that
%
\begin{equation}
\label{DefineOmega} \Omega(i,j) = \theta(i) \theta(j) P(k, \ell)\qquad\mbox{if } i
\in V^{(k)} \mbox{ and }j \in V^{(\ell)}.
\end{equation}
Similarly,
for identifiability, we fix a constant $g_0 \in(0,1)$ and assume that
%
\begin{equation}
\label{DefineA} \max_{1 \leq i, j \leq K} P(i,j) = 1, \qquad0< \thmin \leq
\thmax\leq g_0,
\end{equation}
where $\thmin= \min_{1 \leq i \leq n} \{\theta(i)\}$ and $\thmax=
\max_{1 \leq i \leq n} \{ \theta(i)\}$.
Throughout this paper, we assume
%
\begin{equation}
\label{ConditionA} \mbox{$P$ is symmetric, nonsingular, nonnegative and
irreducible}.
\end{equation}
A matrix is nonnegative if all coordinates are nonnegative. See \cite{Horn}, page~361, for the definition of irreducible.

In the analysis below, we use $n$ as the driving asymptotic parameter,
and allow
the vector $\theta$ [and so also the matrix $\Theta$; see (\ref
{DefineTheta})] to depend on $n$.
However, we keep $(K, P)$ as fixed. Consequently, there is a constant
$C = C(P) > 0$ such that
$\llVert P^{-1} \rrVert \leq C$, where $\llVert \cdot\rrVert $
denotes the spectral norm.


The DCBM we use is similar to that in \cite{Zhaoetal} (see also \cite{KarrerNewman2011}), but is different in important ways. In their asymptotic
analysis, Zhao et~al. \cite{Zhaoetal},
model $\theta(i)$ as random variables that have the same means and
take only finite values.
In our setting, we treat $\theta(i)$ as nonrandom and only impose
some mild
regularity conditions and moderate deviations conditions (see below).
Additionally, Zhao et~al. \cite{Zhaoetal} need certain
conditions on $P$ which we do not require. For example,
in the special case of $K = 2$, they require $P$ to be positive
definite, but we do not. See \cite{Zhaoetal}, page~7, for details.

\subsection{Spectral analysis of \texorpdfstring{$\Omega$}{Omega}}
\label{subsecSCOmega}
We start by characterizing the leading eigenvalues and eigenvectors of
$\Omega$.
Recall that
\[
\theta= \bigl(\theta(1), \ldots, \theta(n) \bigr)', \qquad V =
V^{(1)} \cup V^{(2)} \cup\cdots \cup V^{(K)}.
\]
Similarly as before, let $\theta^{(k)}$ be the $n \times1$ vectors
such that
%
\begin{equation}
\label{Definethetak} \theta^{(k)}(i) = \theta(i)\mbox{ or }0,\qquad\mbox{according
to }i \in V^{(k)}\mbox{ or not, }1 \leq k \leq K.
\end{equation}
Let $D$ be the $K \times K$ diagonal matrix of the \textit{overall degree
intensities}
\[
D(k,k) = \bigl\llVert\theta^{(k)} \bigr\rrVert/ \llVert\theta \rrVert,
\qquad1 \leq k \leq K;
\]
note that $D$ depends on $\theta$ and so it also depends on $n$.

The spectral analysis on $\Omega$ hinges on the $K \times K$ matrix
$\DPD$, where $D$ and $P$ are as above.
The following lemma characterizes the
leading eigenvalues and leading eigenvectors of $\Omega$, and is
proved in the supplementary material \cite{score}, Appendix~\textup{C}.


\begin{lemma} \label{lemmabasicalg1}
Suppose\vspace*{1pt} all eigenvalues of $\DPD$ are simple. Let
$\lambda_1/\llVert \theta\rrVert ^2$, $\lambda_2/\llVert \theta
\rrVert ^2, \ldots,
\lambda_K/\llVert \theta\rrVert ^2$ be such eigenvalues, arranged
in the
descending order of the magnitudes, and let
$a_1, a_2, \ldots, a_K$ be the associated (unit-norm) eigenvectors.
Then the $K$ nonzero eigenvalues of $\Omega$ are
$\lambda_1$, $\lambda_2, \ldots, \lambda_K$, with the
associated (unit-norm) eigenvectors being
\[
\eta_k = \sum_{i = 1}^K \bigl[
a_k(i) / \bigl\llVert\theta^{(i)} \bigr\rrVert \bigr] \cdot
\theta^{(i)}, \qquad k = 1, 2, \ldots, K.
\]
\end{lemma}

Note that $(a_k, \eta_k)$ are uniquely determined up to a factor of
$\pm$1; such factors do not affect clustering results.

\subsection{Spectral analysis of $A$}\label{subsecSCX}
In this section, we characterize the leading eigenvalues and leading
eigenvectors of $A$.

Consider the eigenvalues first. The study contains two key components,
one is to characterize
the spectral norm of the noise matrix $W$, and the other is
to impose some conditions on the eigen-spacing of the
matrix $\DPD$ so that the space spanned by the $K$ leading
eigenvectors of $\Omega$ are stable up to noise corruption.

For the first component, recall that $\theta$ may depend on $n$.
We suppose
%
\begin{equation}
\label{Asymp1} \bigl( \log(n) \thmax\llVert\theta\rrVert_1 \bigr) /
\llVert\theta\rrVert^4 \rightarrow0\qquad\mbox{as }n \rightarrow
\infty.
\end{equation}
Combining (\ref{Asymp1}) with basic algebra, it follows that
%
\begin{equation}
\label{Asymp2} \log(n) / \llVert\theta\rrVert^2 \rightarrow0, \qquad
\bigl(\log(n) \llVert\theta\rrVert_1 \llVert\theta\rrVert
_3^3 \bigr) / \llVert\theta\rrVert^6
\rightarrow0,
\end{equation}
which are frequently used in the proofs in the supplementary material \cite{score}.
The following lemma characterizes the spectral norm of $W - \diag(\Omega)$, and is proved in the supplementary material \cite{score}, Appendix~\textup{C}, where
the recent result by \cite{Tropp} on matrix-form Bernstein inequality
is very helpful.

\begin{lemma} \label{lemmaW}
If (\ref{Asymp1}) holds, then
with probability at least $1 + o(n^{-3})$,
\[
\bigl\llVert W - \diag (\Omega) \bigr\rrVert\leq4 \sqrt{\log (n) \thmax \llVert
\theta\rrVert_1}.
\]
\end{lemma}

We wish that the $K$ leading eigenvalues of $A$ are properly spaced and
all of them are
bounded away from $0$. To ensure that, we need some mild conditions on~$\DPD$.
In detail, for any symmetric $K \times K$ matrix $P$,
we denote the minimum gap between
adjacent eigenvalues of $P$ by
%
\begin{equation}
\label{Defineeigsp} \eigsp(P) = \min_{1 \leq i \leq K-1} \llvert
\lambda_{i+1} - \lambda_i\rrvert, \qquad
\lambda_1 > \lambda_2> \cdots>
\lambda_K.
\end{equation}
When any of the eigenvalues of $P$ is not simple, $\eigsp(P) = 0$ by
convention.
We assume that there is a constant $C > 0$ such that for sufficiently
large $n$,
%
\begin{equation}
\label{ConditionDAD} \eigsp(\DPD) \geq C.
\end{equation}
Additionally, we assume the degrees in each communities have
comparable ``overall degree intensities'', in that there is a constant
$h_2 > 0$ such that
%
\begin{equation}
\label{ConditionTheta} \max_{1 \leq i, j \leq K} \bigl\{ \bigl\llVert
\theta^{(i)} \bigr\rrVert/ \bigl\llVert\theta^{(j)} \bigr\rrVert
\bigr\} \leq h_2.
\end{equation}
As a result, $D$ has a bounded condition number.
Recalling that $P$ is a fixed matrix with $\llVert P^{-1}\rrVert \leq C$,
combining this with (\ref{ConditionA}) gives that
all eigenvalues of $\DPD$ are bounded away from either $0$ or $\infty$
by some constants.
Combining these with Lemma~\ref{lemmabasicalg1}, the following lemma
is a direct result of
Lemma~\ref{lemmaW} and basic algebra (e.g., \cite{Bai}, page~473), the
proof of which is omitted.

\begin{lemma} \label{lemmafirsteigen2}
Consider a DCBM where (\ref{Asymp1}), (\ref{ConditionDAD}) and (\ref
{ConditionTheta}) hold. Let $\hlam_1$, $\hlam_2, \ldots, \hlam
_K$ be the leading eigenvalues of $A$, and let $\lambda_1/\llVert
\theta\rrVert
^2$, $\lambda_2/\llVert \theta\rrVert ^2, \ldots,\break  \lambda
_K/\llVert \theta\rrVert ^2$
be the nonzero eigenvalues of $\DPD$, both sorted descendingly in magnitudes.
With probability at least $1 + o(n^{-3})$, the $K$ leading eigenvalues
of $A$ are all simple,
and
\[
\max_{1 \leq k \leq K} \bigl\{ \llvert\hat{\lambda}_k -
\lambda_k \rrvert \bigr\} \leq4 \sqrt{\log(n) \thmax\llVert\theta
\rrVert_1}.
\]
\end{lemma}

Combining Lemma~\ref{lemmafirsteigen2} with (\ref{Asymp1}), with
probability at least $1 + o(n^{-3})$,
%
\begin{equation}
\label{lemmafirsteigen2add} \hat{\lambda}_k \asymp\llVert\theta \rrVert
^2\qquad\mbox{for all }1 \leq k \leq K.
\end{equation}
This result is frequently used in the proof section in the supplementary material~\cite{score}.

Next,\vspace*{2pt} we study the leading eigenvectors.
From now on, we assume conditions (\ref{ConditionDAD})--(\ref{ConditionTheta}) hold,
and let $\hlam_1$, $\hlam_2, \ldots, \hlam_K$ be the $K$
leading eigenvalues as in Lemma~\ref{lemmafirsteigen2}. For $1 \leq k
\leq K$, whenever $\hlam_k$ is not an eigenvalue of
$W - \diag (\Omega)$, let $B^{(k)}$ be the $K \times K$ matrix
\begin{eqnarray}
B^{(k)}(i,j) &=& \bigl( \bigl\llVert\theta^{(i)} \bigr\rrVert
\bigl\llVert\theta^{(j)} \bigr\rrVert \bigr)^{-1} \bigl(
\theta^{(i)} \bigr)' \bigl[I_n - \bigl(W - \diag
(\Omega) \bigr)/\hlam_k \bigr]^{-1} \theta^{(j)},\nonumber
\\
\eqntext{1 \leq i, j \leq K.}
\end{eqnarray}
If $\hlam_k$ is an eigenvalue of $W - \diag (\Omega)$,
let $B^{(k)}$ be the $K \times K$ matrix of $0$.


\begin{lemma} \label{lemmabasicalg2}
Consider a DCBM where (\ref{Asymp1}), (\ref{ConditionDAD}) and (\ref
{ConditionTheta}) hold. Let $\{\hlam_k\}_{k = 1}^K$ be the eigenvalues
of $A$ with the largest magnitudes. There is an event with probability
at least $1 + o(n^{-3})$ such that over the event, for each $1 \leq k
\leq K$, $\hlam_k$ is simple, and the associated eigenvector is given by
\[
\hat{\eta}_k = \sum_{\ell= 1}^K
\bigl(\hat{a}_k(\ell)/ \bigl\llVert\theta^{(\ell)} \bigr\rrVert
\bigr) \bigl[I_n - \bigl(W - \diag (\Omega) \bigr) /
\hlam_k \bigr]^{-1} \theta^{(\ell)},
\]
where $\hat{a}_k$ is an (unit-norm) eigenvector of $\DPDB^{(k)}$, and
$\hlam_k/\llVert \theta\rrVert ^2$ is
the unique eigenvalue of $\DPDB^{(k)}$ that is associated with $\hat{a}_k$.
\end{lemma}

We remark that $\heta_k$ do not necessarily have unit norms, and they
are uniquely determined up to a scaling factor. Among them, $\heta_1$
is particularly interesting, where provided that the network ${\mathcal
N} = (V, E)$ is connected, then all entries of $\heta_1$ are
strictly positive (or strictly negative). Also, the associated
eigenvalue $\hlam_1$ is always strictly positive. These results are
due to Perron's powerful theorem \cite{Horn}, page~508; see Section~\ref{subsecstabi} for more discussion.

\subsection{Characterization of the matrix \texorpdfstring{$\hat{R}^*$}{hat{R}*}}
\label{subsecR}
We now characterize the matrix $\hat{R}^*$, defined as in (\ref
{DefinehatR}). Let $\eta_1, \eta_2, \ldots, \eta_K$ be the $K$
leading (unit-norm) eigenvectors of~$\Omega$ as in Lemma~\ref{lemmabasicalg1}. Define an $n \times(K-1)$ matrix $R$ as a
nonstochastic counterpart of $\hat{R}^*$ by
\[
R(i,k) = \eta_{k+1}(i) / \eta_1(i), \qquad1 \leq k \leq
K-1, 1 \leq i \leq n;
\]
note that $\llVert \eta_k\rrVert =1$. Unlike $\hat{R}$,
$\llvert R(i,k)\rrvert \leq C$ for all $i$ and $k$ (see Lemma~\ref{lemmabasicalg1}), so it is
unnecessary to impose thresholding as that in (\ref{DefinehatR}).

We wish to characterize $\llVert \hat{R}^* - R\rrVert _F$, where
$\llVert \cdot\rrVert _F$
denotes the
Frobenius norm.
To do so, we need to characterize $\llVert \heta_k - \eta_k\rrVert $ and
$\llVert \Theta^{-1} (\heta_k - \eta_k)\rrVert $ [the latter is necessary
because in the definition of $R(i,k)$, we have $\eta_1(i)$ on the
denominator, which will be shown to be at the magnitude of $\theta
(i)$; see Section~\ref{subsecstabi} for details].

To derive tight bounds on $\llVert \heta_k - \eta_k\rrVert $ and
$\llVert \Theta^{-1} (\heta_k - \eta_k)\rrVert $, we need the following
regularity condition, which requires that for sufficiently large $n$,
%
\begin{equation}
\label{maintheta} \log(n) \thmax^2 / \thmin\leq\llVert\theta\rrVert
_3^3.
\end{equation}
This condition can be replaced by more relaxed conditions, B.1--B.2, to
be introduced in the supplementary material \cite{score}, Appendix~\textup{B};
see details therein.

Given the above regularity conditions,
we show in Lemmas~\ref{lemmaell2norm1}--\ref{lemmaell2norm2} (to be
introduced in Section~\ref{subsecstabi}) that
\[
\llVert\heta_k - \eta_k\rrVert^2 \ll \bigl
\llVert\Theta^{-1}(\heta_k - \eta_k) \bigr
\rrVert^2, \qquad \bigl\llVert\Theta^{-1}(
\heta_k - \eta_k) \bigr\rrVert^2 \leq C
\log(n) \err_n,
\]
where
%
\begin{equation}
\label{Defineerr} \err_n = \frac{\llVert \theta\rrVert _3^3}{\llVert
\theta\rrVert ^6} \Biggl[ \sum
_{i = 1}^n \frac{1}{\theta(i)} + \frac{\log(n)}{\thmin}
\biggl( \frac{\llVert \theta\rrVert _1}{\llVert \theta\rrVert ^2} \biggr)^2 \Biggr],
\end{equation}
and the right-hand side are bounds derived from Taylor expansions
of\break $[\Theta^{-1}(\heta_k - \eta_k)]$ and Bernstein inequalities on
random matrices; see the supplementary material
\cite{score}, Appendix~C.12, for details.
It is now not surprising that the leading term of $\llVert \hat{R}^* -
R \rrVert
_F^2$ is determined by $\llVert \Theta^{-1}(\heta_k - \eta_k)\rrVert
$ (and so
by $\err_n$).
The following theorem is the corner stone for characterizing the
behavior of SCORE, and is proved in the supplementary material \cite{score}, Appendix~\textup{B}.

\begin{teo} \label{teoR}
Consider a DCBM where the regularity conditions (\ref{Asymp1}), (\ref
{ConditionDAD}), (\ref{ConditionTheta}) and (\ref{maintheta}) hold.
If $T_n = \log(n)$ is as in (\ref{Definetn}), then as $n \rightarrow
\infty$, with probability at least $1 + o(n^{-2})$, we have that
\[
\bigl\llVert\hat{R}^* - R \bigr\rrVert_F^2 \leq C
\log^3(n) \err_n.
\]
\end{teo}

For general choice of $T_n$, the result continues to hold if we replace
the right-hand side
by $C \log(n) T_n^2 \err_n$.

\subsection{Hamming errors of SCORE}\label{subsecHamm}
Recall that $V = V^{(1)} \cup V^{(2)} \cup\cdots\cup V^{(K)}$ is the true
community partition.
Introduce the $n \times1$ vector $\ell$ of true labels such that
\[
\ell(i) = k\quad\mbox{if and only if}\quad i \in V^{(k)},\qquad 1 \leq i \leq n.
\]
For any community detection procedure, there is a (disjoint) partition
$V = \hat{V}^{(1)} \cup\hat{V}^{(2)} \cdots\cup\hat{V}^{(K)}$, so
we can similarly define the $n \times1$ vector of estimated labels by
\[
\hat{\ell}(i) = k\quad\mbox{if and only if}\quad i \in\hat{V}^{(k)},\qquad
1 \leq i \leq n.
\]
Especially, let $\hat{\ell}^{\sc} = \hat{\ell}^{\sc}(A, T_n, n)$ be
the vector of estimated labels by SCORE.

For any $\hat{\ell}$,
the expected number of mismatched labels is
\[
H_p(\hat{\ell}, \ell) = \sum_{i = 1}^n
P \bigl(\hat{\ell}(i) \neq\ell(i) \bigr).
\]
With that being said, we must note that the clustering errors should
not depend on how we label each of the $K$ communities. Toward this
end, let
%
\begin{equation}
\label{DefineSk} S_K = \bigl\{\pi\dvtx \mbox{$\pi$ is a permutation of
the set $\{1, 2, \ldots, K\}$} \bigr\}.
\end{equation}
Also, for any label vector $\ell$ where the coordinates take value
from $\{1, 2, \ldots, K\}$ and any $\pi\in S_K$, let $\pi(\ell)$
denote the $n \times1$ label vector such that
\[
\pi(\ell) (i) = \pi \bigl( \ell(i) \bigr), \qquad1 \leq i \leq n.
\]
With this notation,
a proper way to measure the performance of $\hat{\ell}$ is to use the
Hamming distance as follows:
\[
\hamm_n( \hat{\ell}, \ell) = \min_{\pi\in S_K}
H_p \bigl(\hat{\ell}, \pi(\ell) \bigr).
\]
For $k = 1, 2, \ldots, K$, let $n_k$ be the size of the $k$th community:
\[
n_k = \bigl\llvert V^{(k)} \bigr\rrvert.
\]
The following theorem is proved in the supplementary material \cite{score}, Appendix~\textup{B}, and is the main result of the paper.

\begin{teo} \label{teomain}
Consider a DCBM where both the regularity conditions
(\ref{Asymp1}), (\ref{ConditionDAD}), (\ref{ConditionTheta}) and
(\ref{maintheta})
hold. Suppose as $n \rightarrow\infty$,
\[
\log^3(n) \err_n/ \min\{n_1, n_2,
\ldots, n_K\} \rightarrow0,
\]
where $\err_n$ is as in (\ref{Defineerr}). For the estimated label
vector $\hat{\ell}^{\sc}$ by the SCORE where the threshold $T_n = \log
(n)$ is as in (\ref{Definetn}), there is a constant $C > 0$ such that
for sufficiently large $n$,
\[
\hamm_n \bigl(\hat{\ell}^{\sc}, \ell \bigr) \leq C
\log^3(n) \err_n.
\]
\end{teo}

Similarly, for general $T_n$, the theorem continues to hold if we
replace the right-hand side by
$C \log(n) T_n^2 \err_n$. Theorem~\ref{teomain} says that SCORE is
(weakly) consistent under mild conditions; see \cite{Zhaoetal} for
difference between strong consistency and weak consistency.

\subsection{Stability of SCORE} \label{subsecstabi}
The performance of SCORE hinges on the matrix $\hat{R}$ defined in
(\ref{DefinehatR}):
\[
\hat{R}(i,k) = \heta_{k +1}(i) / \heta_1(i), \qquad1 \leq i
\leq n, 1 \leq k \leq K-1.
\]
Seemingly, SCORE could be
unstable if the denominator $\heta_1(i)$ is small (or even worse,
equals to $0$) for some $i$. Fortunately, this is not the case, and
under mild conditions,
for most $i$ (or for all $i$ with slightly stronger conditions), $\heta
_1(i) \asymp\eta_1(i) \asymp\theta(i)$.
Below, we further characterize the vector $(\heta_k - \eta_k)$, with
emphasis on the case of $k = 1$.

We start with the following lemma, which is the direct result of
Perron's theorem~\cite{Horn}, Section~8.2, on nonnegative matrices,
and which says that
a coordinate of $\heta_1$ can never be exactly $0$, as long as the
network is connected.

\begin{lemma} \label{lemmaheta1}
Let $A$ be the adjacency matrix of a network ${\mathcal N} = (V, E)$,
let $\hlam_1$ be the eigenvalue with the largest magnitude, and let
$\heta_1$ be the associated eigenvector where at least one coordinate
is positive. If ${\mathcal N}$ is connected, then both $\hlam_1$ and
all coordinates of $\heta_1$ are strictly positive.
\end{lemma}

Next, for any $n \times1$ vector $\xi$ with strictly positive
coordinates, define the \textit{coordinate oscillation} by
\[
\OSC(\xi) = \max_{1 \leq i, j \leq n} \bigl\{ \xi(i) / \xi(j) \bigr\}.
\]
The following lemma is proved in the supplementary material \cite{score}, Appendix~\textup{C}
[note that the $i$th coordinate of $\Theta^{-1}
\eta_1$ is $\eta_1(i)/\theta(i)$].

\begin{lemma} \label{lemmaosc}
Consider a DCBM where (\ref{ConditionDAD})--(\ref{ConditionTheta})
holds. We have
\[
\OSC \bigl(\Theta^{-1} \eta_1 \bigr) \leq C.
\]
\end{lemma}

The following lemmas
constitute the key component of the proof of
Theorem~\ref{teoR}, but can also be used to obtain upper bounds for
the number of
``ill-behaved'' coordinates of $\heta_1$.
These lemmas are proved in the supplementary material \cite{score}, Appendix~\textup{C}.

\begin{lemma} \label{lemmaell2norm1}
Consider a DCBM where the conditions of Theorem~\ref{teoR} hold. With
probability at least $1 + o(n^{-3})$, for all $1 \leq k \leq K$,
\[
\llVert\heta_k - \eta_k\rrVert^2 \leq C
\log(n) \llVert\theta\rrVert_1 \llVert\theta\rrVert
_3^3 / \llVert\theta\rrVert^6.
\]
\end{lemma}


\begin{lemma} \label{lemmaell2norm2}
Consider a DCBM where the conditions of Theorem~\ref{teoR} hold.
With probability at least $1 + o(n^{-3})$, for all $1 \leq k \leq K$,
\[
\bigl\llVert\Theta^{-1} (\heta_k - \eta_k)
\bigr\rrVert^2 \leq C \log(n) \err_n.
\]
\end{lemma}

Recall that $\err_n$ is defined in (\ref{Defineerr}). For Lemma~\ref{lemmaell2norm2},
a weaker bound is possible if we simply combine Lemma~\ref{lemmaell2norm1} and the fact that $\llVert \Theta^{-1}\rrVert \leq
1/\thmin$.
The current bound is much sharper, especially when only a few $\theta
(i)$ are small.

We now obtain an upper bound on the number of ``ill-behaved'' entries of
$\heta_1$. Recall that
$\OSC(\Theta^{-1} \eta_1) \leq C$. Fixing a constant $c_0 \in(0,1)$,
we call the $i$th enty of $\heta_1$ well-behaved if $\llvert \heta
_1(i) /
\eta_1(i) - 1\rrvert \leq c_0$ (say). Let
%
\begin{equation}
\label{Defineshat} \quad \hat{S} = \hat{S}(c_0, \heta_1,
\eta_1; A, \Omega, n) = \bigl\{1 \leq i \leq n\dvtx \bigl\llvert
\heta_1(i) / \eta_1(i) - 1 \bigr\rrvert\leq
c_0 \bigr\}.
\end{equation}
The following lemma is a direct result of Lemmas~\ref{lemmaosc}~and~\ref{lemmaell2norm2}, so we omit the proof.

\begin{lemma} \label{lemmawell}
Consider a DCBM where the conditions of Theorem~\ref{teoR} hold. Fix\vspace*{1pt}
$c_0 \in(0,1)$ and let $\hat{S}$ be as in (\ref{Defineshat}). Then
with probability at least $1 + o(n^{-3})$, $\llvert V \setminus\hat
{S}\rrvert
\leq C \log(n) \err_n$.
\end{lemma}

Therefore, as long as $\log(n) \err_n/n \rightarrow0$ when $n
\rightarrow\infty$, the fraction of ``ill-behaved'' coordinates of
$\heta_1$ tends to $0$ and is negligible.

In principle, provided that some stronger conditions are imposed, the
techniques in this paper (especially those in the proof of Lemmas~\ref{lemmaell2norm1}--\ref{lemmaell2norm2}) can be used to show that
with probability at least $1 + o(n^{-3})$,
\[
\max_{1 \leq i \leq n} \biggl\llvert\frac{\heta_1(i)}{\eta_1(i)} -1 \biggr\rrvert
\leq c_0,
\]
where $c_0 \in(0,1)$ is a constant.
In this case, Theorem~\ref{teomain} can be strengthened into that of
with probability at least $1 + o(n^{-2})$,
\[
\hamm_n \bigl(\hat{\ell}^{\sc}, \ell \bigr) = 0.
\]
Using terminology in the literature on variable selection \cite{FanLi},
this says that SCORE has the \textit{oracle property}, means that it
achieves \textit{exact recovery} with overwhelming probabilities.

\subsection{Remarks on the regularity conditions} \label{subseciidmodel}
In the main results, Theorems \mbox{\ref{teoR}--\ref{teomain}}, we have
imposed the following regularity conditions: (\ref{Asymp1}), (\ref
{maintheta}) and (\ref{ConditionDAD})--(\ref{ConditionTheta}). The
last two can be roughly translated
to that the leading eigenvalues of $\DPD$ are well spaced and are well
understood.
For this reason, we only discuss (\ref{Asymp1}) and (\ref{maintheta}).

The seeming complexity of (\ref{Asymp1}) and (\ref{maintheta}) is due
to that
we choose \textit{not} to impose much structural assumptions on $\theta$.
The reasons for doing so are carefully explained in Section~\ref{subsecconsi}. On the other hand, if we choose to impose some
structural assumptions on $\theta$,
these conditions can be much simplified.
In this section, we illustrate this with an (scaled) i.i.d. model for
$\theta$.

In detail, let $F$ be a distribution defined over $(0, \infty)$
and that does not vary with~$n$, and let $g_n > 0$ be a (nonrandom)
scaling factor.
We model
\[
\theta= g_n \cdot\mu\qquad\mbox{where } \mu= \bigl(\mu(1), \ldots,
\mu(n) \bigr)'\mbox{ and } \mu(i) \stackrel{\mathrm{i.i.d.}}
{\sim} F;
\]
we allow $g_n \rightarrow0$ as $n \rightarrow\infty$, but require
$\sqrt{n} g_n \rightarrow\infty$.
For $q \neq0$, let $m_q(F) = \int_0^{\infty} x^{q} \,d F(x)$.
Suppose $F$ satisfies some regularity conditions. By basic statistics,
except for negligible probabilities,
%
\begin{equation}
\label{LLN} \llVert\theta\rrVert_q^q =
g_n^q \llVert\mu\rrVert_q^q \sim
g_n^q m_q(F) n, \qquad q = -1, 1, 2, 3.
\end{equation}
Similarly, let $\mumax$ and $\mumin$ be the largest and smallest
entry of $\mu$, respectively.
Using (\ref{LLN}), (\ref{Asymp1}) and (\ref{maintheta}) are
satisfied if
%
\begin{equation}
\label{MDM} \frac{\log(n) \mumax^2/\mumin}{ n g_n^2} \rightarrow0.
\end{equation}
Also, note that when (\ref{MDM}) holds,
\[
\err_n \leq C \bigl(n g_n^2
\bigr)^{-2}.
\]
Below are some examples where (\ref{MDM}) holds.
%
\begin{itemize}
\item(\textit{Finite support}). The support of $F$ is contained in $[a,
b]$ for constants \mbox{$b > a > 0$}. This includes the Block Model as a
special case (e.g., \cite{Yu,Fishkind}), where $F$ is a point mass at
$1$. In this case, (\ref{MDM}) reduces to $\log(n) / [n g_n^2]
\rightarrow0$. This is similar to that in \cite{Zhaoetal}, page~7, where
$F$ is supported on $M$ different points $x_1, x_2, \ldots, x_M$.
\item($F$ \textit{is log-normal}). A frequently used model is when
$F$ is the CDF of $e^{X}$, where $X \sim N(u, \sigma^2)$. In this
case, except for a probability of $o(n^{-k})$, $\mumin\geq\exp (u - \sigma\sqrt{2 k \log(n)})$ and $\mumax\leq\exp (u + \sigma\sqrt{2 k \log(n)})$, and\break $\mumax^2/\mumin\ll
n^{\delta}$ for any fixed $\delta> 0$ and large $n$.
Therefore, condition (\ref{MDM}) holds as long as $g_n \geq
n^{-\vartheta}$ for some $0 \leq\vartheta< 1/2$.
\item(\textit{Polynomial tails}). The most difficult case is when $F$ has
polynomial tails.
If $F(x) \geq c_1 x^{q_1}$ as $x \rightarrow0$ and $(1 - F(x)) \leq
c_2 x^{-q_2}$ as $x \rightarrow\infty$ for constants $c_1$, $c_2$, $q_1$, $q_2$,
where $q_1$ and $q_2$ are sufficiently large so that (\ref{LLN})
continues to hold.
In this case, except for probability $O(n^{-k})$, $\mumin\geq C
n^{-(k+1)/q_1}$ and $\mumax\leq C n^{-(k+1)/q_2}$, and (\ref{MDM}) holds as long as
$\log(n) \cdot n^{1 - (k+1)(2q_2^{-1} + q_1^{-1})} g_n^2 \rightarrow0$.
The discussion applies to the case where the support of $F$ is
contained in $[a, \infty)$ and that $(1 - F(x)) \leq c_1 \exp (-c_2 x^{c_3})$ as $x \rightarrow\infty$, where $a, c_1, c_2,
c_3$ are positive constants. In such cases, we can view $q_1 = q_2 =
\infty$.
\end{itemize}

In conclusion, under the (scaled) i.i.d. model, Theorem~\ref{teoR} holds if we replace (\ref{Asymp1}) and (\ref{maintheta}) by
(\ref{MDM}), where $\err_n$ is simplified as in $C(n g_n^2)^{-2}$;
these results are very similar to those in
\cite{Zhaoetal}.


\section{Simulations} \label{secSimul}
We have conducted a small-scale simulation study. The goal is to
select a few representative cases to investigate the performances of
the procedures we discussed in the preceding sections.

The simulation includes $6$ different experiments, where we compare $5$
different algorithms: SCORE, oPCA, nPCA, the profile likelihood
approach (PL), Newman's Spectral Modularity method (Newman SM) and
pseudo Likelihood (pseudo).

PL is realized with a tabu algorithm, which needs the input of an
initial label vector. We consider two approaches to setting initial
label vector: we generate the label vector randomly or set it as the
estimated label vector by the SCORE. We consider the first approach in
Experiments \ref{ex1}--\ref{ex5} and both approaches in Experiment~\ref{ex6}. To differentiate
two approaches, PL with the second approach by PL, and PL with the
second approach by PL-SCORE.

For each simulation experiment, we choose integers $n$, $K$, and $\rep$,
representing the size of the network, the number of communities, and
the number of repetitions for simulations, correspondingly.
Fix a $K \times K$ matrix $P$.
We generate an $n \times1$ vector $\ell$ taking values from $\{1, 2,
\ldots, K\}$, representing the vector for community labels, and an $n
\times1$ vector $\theta$ representing the degree
heterogeneities. For $k = 1, 2, \ldots, K$, we let $V^{(k)} = \{1 \leq
i \leq n\dvtx  \ell_i = k\}$, and
let ${\mathbf1}_k$ be the indicator vector of $V^{(k)}$ as before.
Each simulation experiment contains the following steps:
\begin{longlist}[(a)]
\item[(a)] Let $\Theta$ be the $n \times n$ diagonal matrix such that
$\Theta(i,i) = \theta(i)$, $1 \leq i \leq n$. Define the $n \times n$
matrix $\Omega$ by $\Omega= \Theta[\sum_{k, \ell= 1}^K P(k,\ell)
{\mathbf1}_k {\mathbf1}_{\ell}' ] \Theta$.
\item[(b)] Generate a symmetric $n \times n$ matrix $W$ where all
diagonals are $0$,
and for all $1 \leq i < j \leq n$, $W(i,j)$ are independent
centered-Bernoulli with parameters $\Omega(i,j)$. Let $\tilde{A} =
\Omega- \diag (\Omega) + W$, which can be viewed as the
adjacency matrix of a network, say, ${\mathcal N} = (V, E)$.
\item[(c)] Let ${\mathcal N}_0 = (V_0, E_0)$ be the giant component of
${\mathcal N} = (V, E)$. Let $A$ be the adjacency matrix of ${\mathcal
N}_0$, and let $n_0$ be the size of ${\mathcal N}_0$.
\item[(d)] Apply all or a subset of the $5$ aforementioned algorithms
to $A$. Record the Hamming error rates of all methods under investigations.
\item[(e)] For integer $\rep$ mentioned above, repeat (b)--(d) for $\rep$ times.
\end{longlist}
Hamming error rate is defined as the ratio between the Hamming errors
and $n_0$.
In our study, $n_0$ is usually very close to $n$ so we do not report
the exact values.
Also, we set the threshold $T_n$ in (\ref{DefinehatR}) as $\infty$ so
that we do not truncate
any coordinates of $\hat{R}$ as usually none of them is unduly large; setting
$T_n = \log(n)$ gives almost the same results.
We now describe the each experiment in detail.

\begin{ex}\label{ex1}
In this experiment, we investigate
how SCORE, oPCA, nPCA, and PL perform with the
classical stochastic Block Model (BM).
We choose $(n, K, \rep) = (1000, 2, 50)$, $P$ as the $2 \times2$ matrix
with $1$ on the diagonals and $0.5$ on the off-diagonals, and $\theta$
as the
vector where all coordinates are $0.2$. Also, we generate the label
vector $\ell$ randomly by $(\ell_i - 1) \stackrel{\mathrm{i.i.d.}}{\sim}
\operatorname{Bernoulli}(1/2)$.
This is a relatively easy case and all methods perform satisfactory and have
similar error rates. See Table~\ref{tableSimuBM1} for the results.
\begin{table}
\tabcolsep=0pt
\caption{Comparison of mean error rates (Experiment \protect\ref{ex1}). In each cell,
the number in the bracket is the corresponding standard deviation (SD)}\label{tableSimuBM1}
\begin{tabular*}{\tablewidth}{@{\extracolsep{\fill}}@{}lcccc@{}}
\hline
\textbf{Methods} & \textbf{oPCA} & \textbf{nPCA} & \textbf{PL} & \textbf{SCORE}
\\
\hline Mean (SD) & 0.058 (0.009) & 0.055 (0.010) & 0.050 (0.065) & 0.058 (0.009)\\
\hline
\end{tabular*}
\end{table}

%
\begin{table}
\tabcolsep=0pt
\caption{Comparison of mean error rates when there are three
communities (Experiment \protect\ref{ex2}). In each cell, the number in the bracket is
the corresponding standard deviation (SD)}\label{tableThreeclasses}
\begin{tabular*}{\tablewidth}{@{\extracolsep{\fill}}@{}lcccc@{}}
\hline
\textbf{Methods} & \textbf{oPCA} & \textbf{nPCA} & \textbf{PL} & \textbf{SCORE}\\
\hline
Mean (SD) &  0.378  (0.041) &  0.165  (0.084) &  0.0636  (0.123) &  0.0695  (0.004)\\
\hline
\end{tabular*}
\end{table}

%
\begin{table}[b]
\tabcolsep=0pt
\caption{Comparison of error rates [Experiment \protect\ref{ex3}\textup{(a)}]. In each cell, the
number in the bracket is the corresponding standard deviation (SD).
From top to bottom: $\theta(i) = d_0 + (c_0 - d_0) (i/n)$, $\theta(i)
= d_0 + (c_0 - d_0) (i/n)^2$ and $\theta(i) = c_0 1\{i \leq n/2\} +
d_0 \{i > n/2\}$; $(c_0, d_0) = (0.5,0.02)$}\label{tableDifferentTheta}
\begin{tabular*}{\tablewidth}{@{\extracolsep{\fill}}@{}lcccc@{}}
\hline
\textbf{Methods} & \textbf{oPCA} & \textbf{nPCA} & \textbf{PL} & \textbf{SCORE}\\
\hline
Mean (SD) & 0.066 (0.021) & 0.066 (0.107) & 0.042 (0.064) & 0.043 (0.006)\\
& 0.292 (0.014) & 0.431 (0.122) & 0.138 (0.080) & 0.140 (0.010)\\
& 0.254 (0.034) & 0.476 (0.049) & 0.139 (0.074) & 0.130 (0.010)\\
\hline
\end{tabular*}
\end{table}

It is noteworthy that in one of the repetitions, PL fails to converge
and has an error
rate of $49.8\%$. Such outlying cases are observed in most experiments
below; sometimes the fraction of
outlying cases is larger.
\end{ex}

\begin{ex}\label{ex2}
In this example, we compare the performance of
oPCA, nPCA, PL and SCORE (we use a slight variant of SCORE, SCOREq with
$q = 2$; see the supplementary material \cite{score}. The performance
of this variant is generally similar to that of SCORE, but is slightly
better in this experiment) for the case where we have three communities.
We take $(n, K, \rep) = (1500, 3, 25)$, and $P$ as the $3 \times3$
symmetric matrix where we have $1$ on the diagonals,
$P(1,2) = 0.4$, $P(2,3) = 0.4$ and $P(1,3) = 0.05$.
We take $\theta$ as the vector such that
$\theta(i) =0.015 +0.785 \times(i/n)^2$, $1 \leq i \leq n$, and
generate the label vector $\ell$ randomly such that $\ell_i = 1, 2,
3$ with equal probabilities.
The results are reported in Table~\ref{tableThreeclasses}, which
suggest that SCORE
outperforms nPCA and oPCA. The error rates of SCORE and PL
are similar, but SCORE is comparably more stable than PL.
\end{ex}

\begin{ex}\label{ex3}
In this experiment, we investigate how the heterogeneity parameters
affect the performances of oPCA, nPCA, SCORE, PL, and Newman's SM. The
experiment contains two
experiments, Experiments \ref{ex3}(a)--\ref{ex3}(b):\vspace*{6pt}

Experiment \ref{ex3}(a). In this experiment, we take $(n, K, \rep) = (1000, 2, 50)$, $P$ as
the $2 \times2$ matrix that has $1$ on the diagonals and $0.5$ on the
off-diagonals. Also, we generate $\ell$ randomly by $(\ell_i - 1)
\stackrel{\mathrm{i.i.d.}}{\sim} \operatorname{Bernoulli}(1/2)$. Fixing $c_0 = 0.5$
and $d_0 = 0.02$, we
investigate three different choices of $\theta$. In the first one,
$\theta(i)
= d_0 + (c_0 - d_0)(i/n)$, $1 \leq i \leq n$.
In the second one, $\theta(i)
= d_0 + (c_0 - d_0)(i/n)^2$, $1 \leq i \leq n$. In the last one,
$\theta(i) = c_0 1\{i \leq n/2\} + d_0 1\{ i > n/2\}$, $1 \leq i \leq n$.
Note that the heterogeneity effects are mild for the first choice of
$\theta$, but are much more severe in the other two choices.
The results are tabulated in Table~\ref{tableDifferentTheta}.
The error rates of oPCA and nPCA are usually higher than that of PL and
the SCORE.
The average error rates of PL and SCORE are similar, but PL usually has
a much larger standard deviation. The instability
of the PL algorithm is due to that it depends on an initial guess
(generated randomly),
and when the initial guess is ``bad,'' PL may fail to converge to the
true labels.\vspace*{6pt}

%
\begin{figure}

\includegraphics{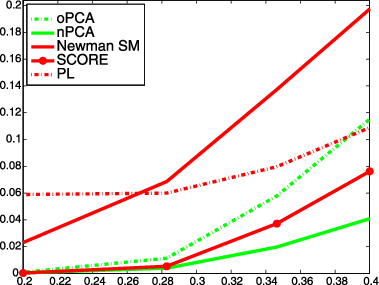}

\caption{Comparison of Hamming error rates [Experiment \protect\ref{ex3}\textup{(b)}]. The
heterogeneity vector $\theta$ satisfies $\log(\theta(i))
\stackrel{\mathrm{i.i.d.}}{\sim} N(0, \sigma^2)$, $1 \leq i \leq n$. $x$-axis:
$\sigma$. $y$-axis: Hamming errors.}\label{figlognorm}
\end{figure}

Experiment \ref{ex3}(b). In this experiment, we take $(n, K, \rep) = (1000, 2, 25)$ and $P$ as the
$2 \times2$ matrix with $1$ on the diagonals and $0.5$ on the
off-diagonals. We set the label vector
$\ell$ by $\ell_i = 1\{i \leq n/4\} + 2 \cdot1\{n/4 < i \leq n \}$.
We first generate $\theta$ by $\log(\theta(i)) \stackrel{\mathrm{i.i.d.}}{\sim
} N(0, \sigma^2)$, $1 \leq i \leq n$, where $\sigma=0.2 \times[1,
\sqrt{2}, \sqrt{3}, 2, \sqrt{5}]$, and then normalize $\theta$ by
$\theta=0.9 \times\theta/\thmax$.
The results are reported in Figure~\ref{figlognorm}, which suggest
that SCORE
has better performance than oPCA, Newman's SM and PL. Somewhat
surprisingly, in this particular setting,
nPCA has the best performance among all these procedures.
\end{ex}

\begin{ex}\label{ex4}
In this experiment,
we study the performances of the procedures
for a larger $n$, and investigate how the heterogeneity
affects the performances of the above procedures.
We take $(n, K, \rep) = (4000, 2, 25)$, $P$ be the symmetric matrix with
$P(1,1) = 3$, $P(2,2) = 1$
and $P(1,2) = 0.5$, and set the label vector $\ell$ by $\ell_i = 1\{i
< n/4\} +
2 \cdot1\{n/4 < i \leq n\}$.
For $c_0 = 0.5$ and each $d_0$ in $0.0025 \times[1, 3, 5, 7, 9]$,
we take
$\theta(1\dvtx n/4) = d_0 + (c_0 - d_0) \times(4i/n)$
and
$\theta(1+ n/4\dvtx n) = d_0 + (c_0 - d_0) \times[4i/(3n)]^2$.
We further normalize two sub-vectors $\theta(1\dvtx n/4) $ and
$\theta(1+n/4\dvtx n)$ by dividing their vector $\ell^1$-norm, respectively.
Finally, the whole vector $\theta$ is normalized by $\theta= 0.8
\times
\theta/ \thmax$.
In this setting,
$\thmax= c_0$ and $\thmin\approx d_0$, so when
$d_0$ decrease, the level of heterogeneity increase.

%
\begin{figure}

\includegraphics{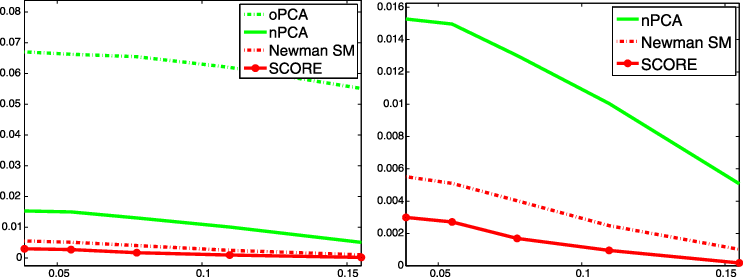}

\caption{Comparison of Hamming error rates (Experiment \protect\ref{ex4}). Right
panel: zoom-in of the left panel. $x$-axis: parameter $d_0$; see
Experiment \protect\ref{ex4} for details. $y$-axis: Hamming errors.}\label{figNewmanSC}
\end{figure}

The performance of oPCA, nPCA, SCORE and Newman's SM are reported in
Figure~\ref{figNewmanSC}. In this experiment, $n = 4000$ and
PL is found to be rather time consuming so
we do not include it in the experiment. The results suggest
that the problem is increasingly harder as $d_0$ decrease,
and SCORE has the best
performance among all $4$ methods.
\end{ex}

\begin{ex}\label{ex5}
In this experiment,
we take $(n, k, \rep) = (1200, 2, 25)$, and~$\theta$ to be the vector with $\theta(i) = d_0 + (c_0-d_0) \times(i/n)^2$,
where $d_0 =0.025$ and $c_0 =0.5$, and investigate the performances
of all the above procedures.
The experiment contains two sub-experiments: Experiment \ref{ex5}(a) and \ref{ex5}(b):\vspace*{6pt}

Experiment \ref{ex5}(a). In this experiment, we investigate how the ratios between the diagonals and
off-diagonals of $P$ (which can be thought of as a measure for how
the classes are separated) affect the performances of all $5$ procedures.
The label vector $\ell$ is randomly generated by $(\ell_i -1)
\stackrel{\mathrm{i.i.d.}}{\sim} \operatorname{Bernoulli}(1/2)$. For each $a$ in $0.1
\times(3,4,5,6,7)$,
we take $P$ as the $2 \times2$ matrix which has $1$ on the diagonals
and $a$ on the off-diagonals.
The results are reported in Figure~\ref{figvaryinga} (left panel), which
suggest that the problems become
increasingly harder as $a$ increase.
In this experiment, SCORE and Newman's SM
have very similar error rates, which are smaller
than those of nPCA, oPCA and PL.\vspace*{6pt}

Experiment \ref{ex5}(b). In this experiment, we investigate how the
probabilities of classes affect the performances of the procedures.
We take $P$ as the $2 \times2$ matrix which has $1$ on the diagonals
and $0.5$ on the off-diagonals. For each $\varepsilon$ in
$0.25 + 0.0625 \times[0, 1, 2, 3, 4]$, we generate the label vector randomly
by $(\ell_i -1) \stackrel{\mathrm{i.i.d.}}{\sim} \operatorname{Bernoulli}(\varepsilon
)$. The
results are in Figure~\ref{figvaryinga} (right panel),
which
suggest that the problems become
easier as $\varepsilon$ increase (so that two communities become
increasingly more balanced), and that SCORE
has the best performance among all $5$ procedures.
\end{ex}

\begin{ex}\label{ex6}
In this experiment,
we compare SCORE with PL and \mbox{PL-SCORE} (note that these
are two versions of PL, where in the first one, the initial label
vector is generated randomly, and in the second one, the initial label
vector is set by the estimate of SCORE). We also include the pseudo
Likelihood algorithm (pseudoL) by Amini et~al. \cite{BickelChen2009} for comparison.

%
\begin{figure}

\includegraphics{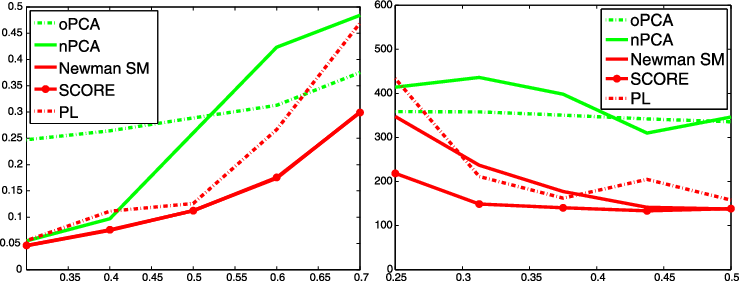}

\caption{Left: Experiment \protect\ref{ex5}\textup{(a)}. $x$-axis: parameter $a$ (the
off-diagonals of $P$); $y$-axis: Hamming error rates in community
detection; the curve for Newman SM is very close to that of SCORE and
is hard to distinguish visually.
Right: Experiment \protect\ref{ex5}\textup{(b)}. $x$-axis: parameter $\varepsilon$ (probability that
a node comes from the community with a smaller size); $y$-axis: Hamming
error rates.}\label{figvaryinga}
\end{figure}

For parameters $(a, c_0, d_0)$ and $\gamma\in\{0.5, 1, 1.5, 2, 2.5\}$,
we take $(n, k, \rep) = (1200, 2, 100)$, let $P$ be the $2 \times2$
matrix which has $1$ on the diagonal and $a$ elsewhere, generate the
label vector $\ell$ by $(\ell_i - 1) \stackrel{\mathrm{i.i.d.}}{\sim} \operatorname{Bernoulli}(1/2)$, and let $\theta(i) = d_0 + (c_0 - d_0) \times
(i/n)^{\gamma}$, $1 \leq i \leq n$. The experiment contains two parts,
Experiment \ref{ex6}(a) and \ref{ex6}(b), where
we take $(a, c_0, d_0)$ to be $(0.8, 0.5, 0.2)$ and $(0.6, 0.5,
0.025)$, respectively.

The results are summarized in Figure~\ref{figgamma}, which suggest
the following. First, SCORE significantly outperforms pseudoL in the
setting of Experiment \ref{ex6}(a) (left panel) and mildly underperforms pseudoL in the
setting of Experiment \ref{ex6}(b) (right panel).
Second, among the four procedures, PL-SCORE has the smallest error
rates. In contrast, the error rates of PL are much larger, which
suggests that for the initial label vector in the tabu algorithm,
letting it be the estimate of the SCORE is substantially better than
letting it be a randomly generated label vector.
However, the improvements of PL-SCORE over SCORE are negligible in the
setting of Experiment \ref{ex6}(a) (in fact, the improvements are negative in some cases)
and are only mild in the setting of Experiment \ref{ex6}(b). The possible reason is that
SCORE may have already provided a good estimate in such settings.
Note that PL-SCORE is computationally more demanding than SCORE,
especially when $n$ is large.

%
\begin{figure}

\includegraphics{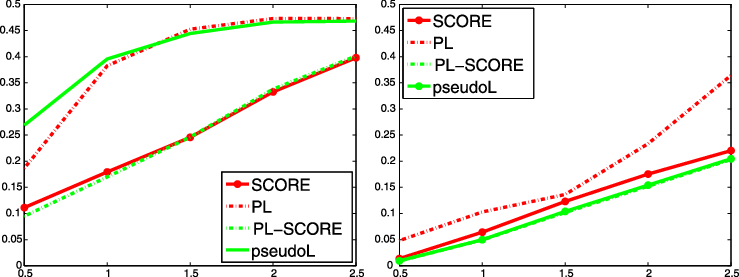}

\caption{Experiment \protect\ref{ex6}\textup{(a)} (left; $a =0.8$) and~\protect\ref{ex6}\textup{(b)} (right; $a =0.6$).
$x$-axis: parameter $\gamma$; $y$-axis: Hamming error rates in
community detection. In Experiment \protect\ref{ex6}\textup{(b)}, the curve associated with
\mbox{PL-SCORE} is very close to that of pseudoL and it is hard to distinguish
them.}\label{figgamma}
\end{figure}

In conclusion, the error rates of SCORE are smaller than those of oPCA,
nPCA, and Newman's SM in many settings. The error rates of SCORE are
similar to those of PL in some settings, but can also
be better in other settings. In the numeric experiments considered
here, we find that PL is computationally slower than oPCA, nPCA and
Newman's SM. It is noteworthy that SCORE is conceptually simple and
easy to implement. This leaves ample space for future work where we can
improve SCORE in various directions, and combine SCORE with other
methods (such as~PL) for better algorithms.
On the other hand, we must note that that SCORE does not always have
the best performance; see, for example, Figure~\ref{figlognorm}. It would
be very interesting to study under which settings SCORE has the best
performance, and under which settings, some other methods presented in
this paper may have a better performance. Seeming, there is no easy
answer to this question, and addressing it needs rather delicate
analysis. For reasons of space, we leave such studies to the future.
\end{ex}

\section{Discussion} \label{secDiscu}
We propose SCORE as a novel spectral approach to
community detection with a DCBM.
The method is largely motivated by the observation that
the degree heterogeneity parameters of the DCBM are largely
ancillary. If we obtain the first $K$ leading eigenvectors
of the adjacency matrix and arrange them in an $n \times K$
matrix $\hat{R}$, then the heterogeneity\vspace*{1pt} can be largely removed by
applying a scaling-invariant mapping to each row of $\hat{R}$.
SCORE is one of such methods.

An interesting feature of SCORE is that it does not attempt to
estimate the heterogeneity parameters or to correct the heterogeneity.
This is especially important when
many nodes of the network are sparse, in which case the estimates
of the heterogeneity parameters are inaccurate and the estimation errors
can largely affect subsequent studies. Additionally, when we tend to
correct the heterogeneity effects, we also tend to inflate the noise
level, resulting a smaller Signal Noise Ratio in spectral analysis.

The theoretic conditions required for the success of the SCORE
is very different from that in Zhao \cite{Zhaoetal}.
Zhao et~al. \cite{Zhaoetal}, page~6, models the
heterogeneity parameters as random variables that assume only finite
values and have the same means, which is relatively restrictive.
For practical concerns, we choose not to put much structural
assumptions on the heterogeneity
parameters, and model them as nonstochastic vectors
that may vary with the size of the network, and we only
need some conditions on regularity and moderate deviations for consistency.
Additionally, Zhao et~al. \cite{Zhaoetal}
impose certain conditions on
the $K \times K$ core matrix $P$ which we do not require.

The work can be extended to various directions. First, SCORE can be
extended to a large class of methods that utilize a scaling-invariant
mapping that operates on $\hat{R}$ row by row. Second, the DCBM
can be generalized to more realistic models, where the spectral
methods could continue to work well. For example, in work in progress
\cite{JinZhang}, we have extended the method to bipartite networks and
have seen nice results on the $110$th Senate and House voting network.
Third, the ideas developed here
can be used to tackle some other problems in network analysis (e.g.,
linkage prediction \cite{Survey}).

In this paper, we have assumed the number of communities $K$ as known.
In many applications (e.g., the web blogs data and the karate club data),
we have a good idea on how many perceivable communities are there, and
such an assumption makes sense. In some other applications (e.g., coexpression
genetic network \cite{Nayak,Liu}), the situation is more complicated
and we may not have
a good idea on how large $K$ is.
Community detection for the case where $K$ is unknown is an unsolved
problem, even for low-dimensional clustering problems.
A possible approach is to try our methods for different
$K$, and see for which $K$ the results give the best fit to the data.
The study along this line is nontrivial and we leave it to the
future work.

We have imposed two regularity conditions in Section~\ref{subsecSCX}:
(\ref{ConditionDAD})--(\ref{ConditionTheta}).
These conditions ensure that the gaps between the leading eigenvalues
of $\Omega$ is lower bounded by $c \llVert \theta\rrVert ^2$ for
some constant
$c > 0$, which in turn are used to
bound the differences between the eigenvalues/eigenvectors of $\Omega$
and those of $A$. On one hand,
such conditions can be relaxed, and the lower bound $c\llVert \theta
\rrVert ^2$
can be replaced by a term of smaller order. On the other hand, when the gap
between two adjacent (nonzero) eigenvalues of $\Omega$
is small (e.g., when the network is sparse \cite{Krza}), it is known
that the associated eigen-vectors are unstable
and are vulnerable to noise corruption; see \cite{Golub}, Section~7.2.5.
In this case, it is usually hard to ensure the stability of the eigenvectors
of $\Omega$, without more assumptions.
The study along this line involves delicate theory on the stability of
eigenvectors \cite{Golub}, and is nontrivial. For this reason, we
leave it for future study.

Intellectually, this work is connected to the recent interest
on low-rank matrix recovery and matrix completion; see, for example,
\cite{Candes}.
In the area of low-rank matrix recovery, there is a tendency
of using the so-called methods of nuclear-norm
penalization to replace spectral clustering.
Our finding says the contrary: spectral clustering can be effective,
and what it takes to make it effective is some careful adjustment.
Such findings are resonated in our forthcoming manuscript
\cite{JinWang}, where we show that spectral clustering can be very effective
in cancer clustering with micro-array data provided that we add a careful
feature selection step.
In spirit, this is connected to several recent papers by Boots and Gordon;
see, for example,~\cite{Gordon}.

\section*{Acknowledgements}
The author would like to thank Joel Tropp and Roman Vershynin for
helpful pointers,
and Stephen E. Fienberg and Peter G. Hall for encouragement.

\begin{supplement}[id=suppA]
\stitle{Supplementary material for ``Fast communication detetion by SCORE''}
\slink[doi]{10.1214/14-AOS1265SUPP} 
\sdatatype{.pdf}
\sfilename{aos1265\_supp.pdf}
\sdescription{Owing to space constraints, the technical proofs are
relegated a supplementary document. The supplementary document contains
three parts: Appendices A, B and C. Appendix A discusses possible
variants of SCORE, Appendix B contains proofs of the main theorems and
Appendix~C contains proofs of the secondary lemmas.}
\end{supplement}

%

\printaddresses

\begin{thebibliography}{38}

\bibitem{AdamicGlance2005}
%
\begin{bincollection}[auto:STB|2014/08/04|07:23:14]
\bauthor{\bsnm{Adamic},~\bfnm{L.}\binits{L.}} \AND
\bauthor{\bsnm{Glance},~\bfnm{N.}\binits{N.}}
(\byear{2005}).
\btitle{The political blogosphere and the 2004 U.S. election: Divided they blog}.
In \bbooktitle{Proceedings of the 3rd International Workshop on Link Discovery}
\bpages{36--43}.
\bpublisher{ACM}, \blocation{New York}.
\end{bincollection}
%
\bptok{imsref}%
\endbibitem

\bibitem{Chenetal}
%
\begin{barticle}[mr]
\bauthor{\bsnm{Amini},~\bfnm{Arash~A.}\binits{A.~A.}},
\bauthor{\bsnm{Chen},~\bfnm{Aiyou}\binits{A.}},
\bauthor{\bsnm{Bickel},~\bfnm{Peter~J.}\binits{P.~J.}} \AND
\bauthor{\bsnm{Levina},~\bfnm{Elizaveta}\binits{E.}}
(\byear{2013}).
\btitle{Pseudo-likelihood methods for community detection in large
sparse networks}.
\bjournal{Ann. Statist.}
\bvolume{41}
\bpages{2097--2122}.
\bid{doi={10.1214/13-AOS1138}, issn={0090-5364}, mr={3127859}}
\end{barticle}
%
\bptok{imsref}%
\endbibitem

\bibitem{Bai}
%
\begin{bbook}[mr]
\bauthor{\bsnm{Bai},~\bfnm{Zhidong}\binits{Z.}} \AND
\bauthor{\bsnm{Silverstein},~\bfnm{Jack~W.}\binits{J.~W.}}
(\byear{2009}).
\btitle{Spectral Analysis of Large Dimensional Random Matrices},
\bedition{2nd} ed.
\bpublisher{Springer},
\blocation{New York}.
\bid{doi={10.1007/978-1-4419-0661-8}, mr={2567175}}
\bptnote{check year}%
\end{bbook}
%
\bptok{imsref}%
\endbibitem

\bibitem{BickelChen2009}
%
\begin{barticle}[pbm]
\bauthor{\bsnm{Bickel},~\bfnm{Peter~J.}\binits{P.~J.}} \AND
\bauthor{\bsnm{Chen},~\bfnm{Aiyou}\binits{A.}}
(\byear{2009}).
\btitle{A nonparametric view of network models and Newman-Girvan and
other modularities}.
\bjournal{Proc. Natl. Acad. Sci. USA}
\bvolume{106}
\bpages{21068--21073}.
\bid{doi={10.1073/pnas.0907096106}, issn={1091-6490},
pii={0907096106}, pmcid={2795514}, pmid={19934050}}
\end{barticle}
%
\bptok{imsref}%
\endbibitem

\bibitem{Gordon}
%
\begin{bincollection}[auto:STB|2014/08/04|07:23:14]
\bauthor{\bsnm{Boots},~\bfnm{B.}\binits{B.}} \AND
\bauthor{\bsnm{Gordon},~\bfnm{G.}\binits{G.}}
(\byear{2011}).
\btitle{Online spectral identification of dynamical systems}.
In \bbooktitle{NIPS Workshop on Sparse Representation and Low-Rank
Approximation}.
\blocation{Sierra Nevada, Spain}.
\end{bincollection}
%
\bptok{imsref}%
\endbibitem

\bibitem{Box}
%
\begin{bbook}[mr]
\bauthor{\bsnm{Box},~\bfnm{George~E.~P.}\binits{G.~E.~P.}} \AND
\bauthor{\bsnm{Draper},~\bfnm{Norman~R.}\binits{N.~R.}}
(\byear{1987}).
\btitle{Empirical Model-Building and Response Surfaces}.
\bpublisher{Wiley},
\blocation{New York}.
\bid{mr={0861118}}
\end{bbook}
\bptok{imsref}%
\endbibitem

\bibitem{Candes}
%
\begin{barticle}[mr]
\bauthor{\bsnm{Cand{\`e}s},~\bfnm{Emmanuel~J.}\binits{E.~J.}},
\bauthor{\bsnm{Li},~\bfnm{Xiaodong}\binits{X.}},
\bauthor{\bsnm{Ma},~\bfnm{Yi}\binits{Y.}} \AND
\bauthor{\bsnm{Wright},~\bfnm{John}\binits{J.}}
(\byear{2011}).
\btitle{Robust principal component analysis?}
\bjournal{J. ACM}
\bvolume{58}
\bpages{Art. 11, 37}.
\bid{doi={10.1145/1970392.1970395}, issn={0004-5411}, mr={2811000}}
\end{barticle}
%
\bptok{imsref}%
\endbibitem

\bibitem{FanLu}
%
\begin{barticle}[auto:STB|2014/08/04|07:23:14]
\bauthor{\bsnm{Chaudhuri},~\bfnm{K.}\binits{K.}},
\bauthor{\bsnm{Fan},~\bfnm{C.}\binits{C.}} \AND
\bauthor{\bsnm{Tsiatas},~\bfnm{A.}\binits{A.}}
(\byear{2012}).
\btitle{Spectral clustering of graphs with general degrees in the
extended planted partition of model}.
\bjournal{J. Mach. Learn. Res.}
\bvolume{35}
\bpages{1--23}.
\end{barticle}
%
\bptok{imsref}%
\endbibitem

\bibitem{Choi}
%
\begin{barticle}[mr]
\bauthor{\bsnm{Choi},~\bfnm{D.~S.}\binits{D.~S.}},
\bauthor{\bsnm{Wolfe},~\bfnm{P.~J.}\binits{P.~J.}} \AND
\bauthor{\bsnm{Airoldi},~\bfnm{E.~M.}\binits{E.~M.}}
(\byear{2012}).
\btitle{Stochastic blockmodels with a growing number of classes}.
\bjournal{Biometrika}
\bvolume{99}
\bpages{273--284}.
\bid{doi={10.1093/biomet/asr053}, issn={0006-3444}, mr={2931253}}
\end{barticle}
%
\bptok{imsref}%
\endbibitem

\bibitem{Chung}
%
\begin{bbook}[mr]
\bauthor{\bsnm{Chung},~\bfnm{Fan~R.~K.}\binits{F.~R.~K.}}
(\byear{1997}).
\btitle{Spectral Graph Theory},
\bedition{1st} ed.
\bseries{CBMS Regional Conference Series in Mathematics}
\bvolume{92}.
\bpublisher{AMS},
\blocation{Providence, RI}.
\bid{mr={1421568}}
\end{bbook}
%
\bptok{imsref}%
\endbibitem

\bibitem{GW}
%
\begin{barticle}[mr]
\bauthor{\bsnm{Erd{\H{o}}s},~\bfnm{L{\'a}szl{\'o}}\binits{L.}},
\bauthor{\bsnm{Yau},~\bfnm{Horng-Tzer}\binits{H.-T.}} \AND
\bauthor{\bsnm{Yin},~\bfnm{Jun}\binits{J.}}
(\byear{2012}).
\btitle{Bulk universality for generalized {W}igner matrices}.
\bjournal{Probab. Theory Related Fields}
\bvolume{154}
\bpages{341--407}.
\bid{doi={10.1007/s00440-011-0390-3}, issn={0178-8051}, mr={2981427}}
\end{barticle}
%
\bptok{imsref}%
\endbibitem

\bibitem{FanLi}
%
\begin{barticle}[mr]
\bauthor{\bsnm{Fan},~\bfnm{Jianqing}\binits{J.}} \AND
\bauthor{\bsnm{Li},~\bfnm{Runze}\binits{R.}}
(\byear{2001}).
\btitle{Variable selection via nonconcave penalized likelihood and its
oracle properties}.
\bjournal{J. Amer. Statist. Assoc.}
\bvolume{96}
\bpages{1348--1360}.
\bid{doi={10.1198/016214501753382273}, issn={0162-1459}, mr={1946581}}
\end{barticle}
%
\bptok{imsref}%
\endbibitem

\bibitem{Fishkind}
%
\begin{bmisc}[auto:STB|2014/08/04|07:23:14]
\bauthor{\bsnm{Fishkind},~\bfnm{D.}\binits{D.}},
\bauthor{\bsnm{Sussman},~\bfnm{D.}\binits{D.}},
\bauthor{\bsnm{Tang},~\bfnm{M.}\binits{M.}} \AND
\bauthor{\bsnm{Vogelstein},~\bfnm{J.}\binits{J.}}
(\byear{2012}).
\bhowpublished{Consistent adjacency-spectral clustering partitioning
for the stochastic model when the model parameters are unknown.
Available at \arxivurl{arXiv:1205.0309}.}
\end{bmisc}
%
\bptok{imsref}%
\endbibitem


\bibitem{Survey}
%
\begin{barticle}[auto:STB|2014/08/04|07:23:14]
\bauthor{\bsnm{Goldenberg},~\bfnm{A.}\binits{A.}},
\bauthor{\bsnm{Zheng},~\bfnm{A.}\binits{A.}},
\bauthor{\bsnm{Fienberg},~\bfnm{S.}\binits{S.}} \AND
\bauthor{\bsnm{Airoldi},~\bfnm{E.}\binits{E.}}
(\byear{2009}).
\btitle{A survey of statistical network models}.
\bjournal{Faund. Trends Mach. Learn.}
\bvolume{2}
\bpages{129--233}.
\end{barticle}
%
\bptok{imsref}%
\endbibitem

\bibitem{Golub}
%
\begin{bbook}[mr]
\bauthor{\bsnm{Golub},~\bfnm{Gene~H.}\binits{G.~H.}} \AND
\bauthor{\bsnm{Van Loan},~\bfnm{Charles~F.}\binits{C.~F.}}
(\byear{1996}).
\btitle{Matrix Computations},
\bedition{3rd} ed.
\bpublisher{Johns Hopkins Univ. Press},
\blocation{Baltimore, MD}.
\bid{mr={1417720}}
\end{bbook}
%
\bptok{imsref}%
\endbibitem

\bibitem{Tibsbook}
%
\begin{bbook}[mr]
\bauthor{\bsnm{Hastie},~\bfnm{Trevor}\binits{T.}},
\bauthor{\bsnm{Tibshirani},~\bfnm{Robert}\binits{R.}} \AND
\bauthor{\bsnm{Friedman},~\bfnm{Jerome}\binits{J.}}
(\byear{2001}).
\btitle{The Elements of Statistical Learning}.
\bpublisher{Springer},
\blocation{New York}.
\bid{doi={10.1007/978-0-387-21606-5}, mr={1851606}}
\end{bbook}
\bptok{imsref}%
\endbibitem

\bibitem{Hoff}
%
\begin{bincollection}[auto:STB|2014/08/04|07:23:14]
\bauthor{\bsnm{Hoff},~\bfnm{P.}\binits{P.}}
(\byear{2007}).
\btitle{Modeling homophily and stochastic equivalence in symmetric
relational data}.
In \bbooktitle{Advances in Neural Information Processing Systems}.
\blocation{Cambridge}.
\end{bincollection}
%
\bptok{imsref}%
\endbibitem

\bibitem{Horn}
%
\begin{bbook}[mr]
\bauthor{\bsnm{Horn},~\bfnm{Roger~A.}\binits{R.~A.}} \AND
\bauthor{\bsnm{Johnson},~\bfnm{Charles~R.}\binits{C.~R.}}
(\byear{1985}).
\btitle{Matrix Analysis}.
\bpublisher{Cambridge Univ. Press},
\blocation{Cambridge}.
\bid{doi={10.1017/CBO9780511810817}, mr={0832183}}
\end{bbook}
%
\bptok{imsref}%
\endbibitem

\bibitem{score}
%
\begin{bmisc}[author]
{\bauthor{\bsnm{Jin},~\binits{J.}}}
(\byear{2014}).
\bhowpublished{Supplement to ``Fast community detection by SCORE''.
DOI:\doiurl{10.1214/14-AOS1265SUPP}}.
\bptok{imsref}%
\end{bmisc}
%
\endbibitem
%
\bptok{imsref}%
\endbibitem

\bibitem{JinWang}
%
\begin{bmisc}[auto:STB|2014/08/04|07:23:14]
\bauthor{\bsnm{Jin},~\bfnm{J.}\binits{J.}} \AND
\bauthor{\bsnm{Wang},~\bfnm{W.}\binits{W.}}
(\byear{2012}).
\bhowpublished{Optimal spectral clustering by Higher Criticism thresholding.
Working manuscript.}
\end{bmisc}
\bptok{imsref}%
\endbibitem

\bibitem{JinZhang}
%
\begin{bmisc}[auto:STB|2014/08/04|07:23:14]
\bauthor{\bsnm{Jin},~\bfnm{J.}\binits{J.}} \AND
\bauthor{\bsnm{Zhang},~\bfnm{Q.}\binits{Q.}}
(\byear{2012}).
\bhowpublished{New spectral methods for community detection with
bipartite networks.
Working manuscript.}
\end{bmisc}
%
\bptok{imsref}%
\endbibitem

\bibitem{KarrerNewman2011}
%
\begin{barticle}[mr]
\bauthor{\bsnm{Karrer},~\bfnm{Brian}\binits{B.}} \AND
\bauthor{\bsnm{Newman},~\bfnm{M.~E.~J.}\binits{M.~E.~J.}}
(\byear{2011}).
\btitle{Stochastic blockmodels and community structure in networks}.
\bjournal{Phys. Rev. E (3)}
\bvolume{83}
\bpages{016107, 10}.
\bid{doi={10.1103/PhysRevE.83.016107}, issn={1539-3755}, mr={2788206}}
\end{barticle}
%
\bptok{imsref}%
\endbibitem

\bibitem{Kolaczyk}
%
\begin{bbook}[mr]
\bauthor{\bsnm{Kolaczyk},~\bfnm{Eric~D.}\binits{E.~D.}}
(\byear{2009}).
\btitle{Statistical Analysis of Network Data: Methods and Models}.
\bpublisher{Springer},
\blocation{New York}.
\bid{doi={10.1007/978-0-387-88146-1}, mr={2724362}}
\end{bbook}
\bptok{imsref}%
\endbibitem

\bibitem{Krza}
%
\begin{bmisc}[auto:STB|2014/08/04|07:23:14]
\bauthor{\bsnm{Krzakala},~\bfnm{F.}\binits{F.}},
\bauthor{\bsnm{Moore},~\bfnm{C.}\binits{C.}},
\bauthor{\bsnm{Mossel},~\bfnm{E.}\binits{E.}},
\bauthor{\bsnm{Neeman},~\bfnm{J.}\binits{J.}},
\bauthor{\bsnm{Sly},~\bfnm{A.}\binits{A.}},
\bauthor{\bsnm{Zdeborova},~\bfnm{L.}\binits{L.}} \AND
\bauthor{\bsnm{Zhang},~\bfnm{P.}\binits{P.}}
(\byear{2012}).
\bhowpublished{Spectral redemption: Clustering sparse networks.
Available at \arxivurl{arXiv:1306.5550}.}
\end{bmisc}
%
\bptok{imsref}%
\endbibitem

\bibitem{Liu}
%
\begin{barticle}[mr]
\bauthor{\bsnm{Liu},~\bfnm{Han}\binits{H.}},
\bauthor{\bsnm{Xu},~\bfnm{Min}\binits{M.}},
\bauthor{\bsnm{Gu},~\bfnm{Haijie}\binits{H.}},
\bauthor{\bsnm{Gupta},~\bfnm{Anupam}\binits{A.}},
\bauthor{\bsnm{Lafferty},~\bfnm{John}\binits{J.}} \AND
\bauthor{\bsnm{Wasserman},~\bfnm{Larry}\binits{L.}}
(\byear{2011}).
\btitle{Forest density estimation}.
\bjournal{J. Mach. Learn. Res.}
\bvolume{12}
\bpages{907--951}.
\bid{doi={10.1016/j.micres.2009.11.010}, issn={1532-4435}, mr={2786914}}
\end{barticle}
%
\bptok{imsref}%
\endbibitem

\bibitem{Nayak}
%
\begin{barticle}[auto:STB|2014/08/04|07:23:14]
\bauthor{\bsnm{Nayak},~\bfnm{R.}\binits{R.}},
\bauthor{\bsnm{Kearns},~\bfnm{M.}\binits{M.}},
\bauthor{\bsnm{Spielman},~\bfnm{R.}\binits{R.}} \AND
\bauthor{\bsnm{Cheung},~\bfnm{V.}\binits{V.}}
(\byear{2009}).
\btitle{Coexpression network based on natural variation in human gene
expression reveals gene interactions and functions}.
\bjournal{Genome Res.}
\bvolume{19}
\bpages{1953--1962}.
\end{barticle}
\bptok{imsref}%
\endbibitem

\bibitem{Newman1}
%
\begin{barticle}[mr]
\bauthor{\bsnm{Newman},~\bfnm{M.~E.~J.}\binits{M.~E.~J.}}
(\byear{2006}).
\btitle{Finding community structure in networks using the eigenvectors
of matrices}.
\bjournal{Phys. Rev. E (3)}
\bvolume{74}
\bpages{036104, 19}.
\bid{doi={10.1103/PhysRevE.74.036104}, issn={1539-3755}, mr={2282139}}
\end{barticle}
%
\bptok{imsref}%
\endbibitem

\bibitem{Newman2}
%
\begin{barticle}[pbm]
\bauthor{\bsnm{Newman},~\bfnm{M.~E~J.}\binits{M.~E.~J.}}
(\byear{2006}).
\btitle{Modularity and community structure in networks}.
\bjournal{Proc. Natl. Acad. Sci. USA}
\bvolume{103}
\bpages{8577--8582}.
\bid{doi={10.1073/pnas.0601602103}, issn={0027-8424},
pii={0601602103}, pmcid={1482622}, pmid={16723398}}
\end{barticle}
%
\bptok{imsref}%
\endbibitem

\bibitem{Wolfe}
%
\begin{bmisc}[auto:STB|2014/08/04|07:23:14]
\bauthor{\bsnm{Perry},~\bfnm{P.}\binits{P.}} \AND
\bauthor{\bsnm{Wolfe},~\bfnm{P.}\binits{P.}}
(\byear{2012}).
\bhowpublished{Null models for network data.
Available at \arxivurl{arXiv:1201.5871}.}
\end{bmisc}
%
\bptok{imsref}%
\endbibitem

\bibitem{eigenspoke}
%
\begin{bincollection}[auto:STB|2014/08/04|07:23:14]
\bauthor{\bsnm{Prakah},~\bfnm{B.~A.}\binits{B.~A.}},
\bauthor{\bsnm{Sridharan},~\bfnm{A.}\binits{A.}},
\bauthor{\bsnm{Seshadri},~\bfnm{M.}\binits{M.}},
\bauthor{\bsnm{Machiraju},~\bfnm{S.}\binits{S.}} \AND
\bauthor{\bsnm{Faloutsos},~\bfnm{C.}\binits{C.}}
(\byear{2010}).
\btitle{Eigenspokes: Surprising patterns and scalable community chipping in large graphs}.
In \bbooktitle{Advances in Knowledge Discovery and Data Mining}
\bpages{435--448}.
\bpublisher{Springer}, \blocation{Berlin}.
\end{bincollection}
%
\bptok{imsref}%
\endbibitem


\bibitem{Yu}
%
\begin{barticle}[mr]
\bauthor{\bsnm{Rohe},~\bfnm{Karl}\binits{K.}},
\bauthor{\bsnm{Chatterjee},~\bfnm{Sourav}\binits{S.}} \AND
\bauthor{\bsnm{Yu},~\bfnm{Bin}\binits{B.}}
(\byear{2011}).
\btitle{Spectral clustering and the high-dimensional stochastic blockmodel}.
\bjournal{Ann. Statist.}
\bvolume{39}
\bpages{1878--1915}.
\bid{doi={10.1214/11-AOS887}, issn={0090-5364}, mr={2893856}}
\end{barticle}
%
\bptok{imsref}%
\endbibitem

\bibitem{Tropp}
%
\begin{barticle}[mr]
\bauthor{\bsnm{Tropp},~\bfnm{Joel~A.}\binits{J.~A.}}
(\byear{2012}).
\btitle{User-friendly tail bounds for sums of random matrices}.
\bjournal{Found. Comput. Math.}
\bvolume{12}
\bpages{389--434}.
\bid{doi={10.1007/s10208-011-9099-z}, issn={1615-3375}, mr={2946459}}
\end{barticle}
%
\bptok{imsref}%
\endbibitem

\bibitem{Tukey}
%
\begin{barticle}[mr]
\bauthor{\bsnm{Tukey},~\bfnm{John~W.}\binits{J.~W.}}
(\byear{1965}).
\btitle{Which part of the sample contains the information?}
\bjournal{Proc. Natl. Acad. Sci. USA}
\bvolume{53}
\bpages{127--134}.
\bid{issn={0027-8424}, mr={0172387}}
\end{barticle}
%
\bptok{imsref}%
\endbibitem

\bibitem{Yan}
%
\begin{barticle}[auto:STB|2014/08/04|07:23:14]
\bauthor{\bsnm{Yan},~\bfnm{X.}\binits{X.}},
\bauthor{\bsnm{Jensen},~\bfnm{J.}\binits{J.}},
\bauthor{\bsnm{Krzakala},~\bfnm{F.}\binits{F.}}
\betal{et~al.}
(\byear{2014}).
\btitle{Model selection for degree-corrected block model}.
\bjournal{J.~Stat. Mech. Theor. Exp.}
\bvolume{2014}
\bpages{P05007}.
\end{barticle}
%
\bptok{imsref}%
\endbibitem

\bibitem{Zachary}
%
\begin{barticle}[auto:STB|2014/08/04|07:23:14]
\bauthor{\bsnm{Zachary},~\bfnm{W.}\binits{W.}}
(\byear{1977}).
\btitle{An information flow model for conflict and fission in small groups}.
\bjournal{J.~Anthropo. Res.}
\bvolume{33}
\bpages{452--473}.
\end{barticle}
%
\bptok{imsref}%
\endbibitem

\bibitem{Zhao}
%
\begin{barticle}[auto:STB|2014/08/04|07:23:14]
\bauthor{\bsnm{Zhang},~\bfnm{S.}\binits{S.}} \AND
\bauthor{\bsnm{Zhao},~\bfnm{H.}\binits{H.}}
(\byear{2012}).
\btitle{Community identification in networks with unbalanced structure}.
\bjournal{Phys. Rev. E}
\bvolume{85}
\bpages{066114}.
\end{barticle}
%
\bptok{imsref}%
\endbibitem

\bibitem{Zhaoetal}
%
\begin{bmisc}[auto:STB|2014/08/04|07:23:14]
\bauthor{\bsnm{Zhao},~\bfnm{Y.}\binits{Y.}},
\bauthor{\bsnm{Levina},~\bfnm{L.}\binits{L.}} \AND
\bauthor{\bsnm{Zhu},~\bfnm{J.}\binits{J.}}
(\byear{2011}).
\bhowpublished{Consistency of community detection in network under
degree-corrected stochastic block models. Available at \arxivurl
{arXiv:1110.3854v3}.}
\end{bmisc}
%
\bptok{imsref}%
\endbibitem
\end{thebibliography}
\end{document}